\documentclass{aa}

\usepackage{subcaption}
\usepackage{gensymb}
\usepackage[varg]{txfonts}
\usepackage[colorlinks=true, linkcolor=blue, urlcolor=blue, citecolor=blue]{hyperref}
\usepackage{xcolor}
\usepackage{siunitx}

\definecolor{darkgreen}{rgb}{0.0, 0.7, 0.0}
\usepackage{euclid}

\newcommand*{\Halpha}{${\rm H}\alpha$ }
\newcommand*{\scipic}{\texttt{SciPIC}}
\newcommand*{\holcon}{\texttt{HOLCon} }
\newcommand*{\dask}{\texttt{Dask} }
\title{Halo Lightcones with Optimised Orientation and Interpolation in Cosmological Simulations
}
\subtitle{an application to mock \Halpha selected galaxies}
\titlerunning{Halo Lightcones and H$\alpha$ Mocks}
\authorrunning{S. Ramakrishnan et al.}
\author{Sujatha Ramakrishnan \inst{\ref{inst1}\thanks{Corresponding author: sramakrishnan@ice.csic.es}}
\and Francisco J. Castander  \inst{\ref{inst1},\ref{inst3}}
\and Elizabeth J. Gonzalez 
\inst{\ref{inst4},\ref{inst5},\ref{inst8}}
\and Martin Eriksen
\inst{\ref{inst4},\ref{inst8}}
\and Zahra Baghkhani
\inst{\ref{inst1},\ref{inst3}}
\and Pablo Fosalba
\inst{\ref{inst1},\ref{inst3}}
\and Jorge Carretero \inst{\ref{inst6},\ref{inst8}}
\and Gabriele Parimbelli \inst{\ref{inst1},\ref{inst7}}
\and Pau Tallada-Crespí
\inst{\ref{inst6},\ref{inst8}}}
\institute{Institute of Space Sciences (ICE, CSIC), E-08193 Barcelona, Spain \label{inst1}
\and Institut d’Estudis Espacials de Catalunya (IEEC), Edifici RDIT, Campus UPC, 08860 Castelldefels (Barcelona), Spain \label{inst3} \and Institut de F\'{\i}sica d’Altes Energies (IFAE), The Barcelona Institute of Science and Technology, Campus UAB, 08193 Bellaterra (Barcelona), Spain  \label{inst4} \and Instituto de Astronomía Teórica y Experimental (IATE-CONICET), Laprida 854, X5000BGR, C\'ordoba, Argentina \label{inst5} \and Centro de Investigaciones Energ\'{e}ticas, Medioambientales y Tecnol\'{o}gicas (CIEMAT), Avenida Complutense 40, E-28040 Madrid, Spain \label{inst6} \and SISSA - International School for Advanced Studies, Via Bonomea 265, 34136 Trieste, Italy
\label{inst7} \and Port d'Informaci\'{o} Cient\'{i}fica (PIC), Campus UAB, C. Albareda s/n, 08193 Bellaterra (Barcelona), Spain \label{inst8}}
\usepackage{natbib,twoopt}
\usepackage{amsmath}
\usepackage{amsfonts} % For math symbols
\usepackage{amssymb}  % For math symbols (like \checkmark if needed, not used here)
\usepackage{booktabs} % For professional-looking tables (e.g., \toprule, \midrule, \bottomrule)

\usepackage{xcolor}
\usepackage{cleveref}
\crefname{section}{Sect.}{Sects.}
\Crefname{section}{Section}{Sections}

\crefname{figure}{Fig.}{Figs.}
\Crefname{figure}{Figure}{Figures}

\crefname{equation}{Eq.}{Eqs.}
\Crefname{equation}{Equation}{Equations}

\crefname{table}{Table}{Tables}
\Crefname{table}{Table}{Tables}

\crefname{appendix}{Appendix}{Appendices}
\Crefname{appendix}{Appendix}{Appendices}

\bibpunct{(}{)}{;}{a}{}{,}             %% natbib format for A&A and ApJ
\makeatletter
  \newcommandtwoopt{\citeads}[3][][]{\href{http://adsabs.harvard.edu/abs/#3}%
    {\def\hyper@linkstart##1##2{}%
     \let\hyper@linkend\@empty\citealp[#1][#2]{#3}}}
  \newcommandtwoopt{\citepads}[3][][]{\href{http://adsabs.harvard.edu/abs/#3}%
    {\def\hyper@linkstart##1##2{}%
     \let\hyper@linkend\@empty\citep[#1][#2]{#3}}}
  \newcommandtwoopt{\citetads}[3][][]{\href{http://adsabs.harvard.edu/abs/#3}%
    {\def\hyper@linkstart##1##2{}%
     \let\hyper@linkend\@empty\citet[#1][#2]{#3}}}
  \newcommandtwoopt{\citeyearads}[3][][]%
    {\href{http://adsabs.harvard.edu/abs/#3}
    {\def\hyper@linkstart##1##2{}%
     \let\hyper@linkend\@empty\citeyear[#1][#2]{#3}}}
\makeatother

\abstract {A critical step to create realistic mock catalogs that support large-scale photometric and spectroscopic sky
surveys is the production of cosmological simulations that accurately model the survey observables taking into account the redshift-dependent galaxy formation and evolution processes. Here we develop an efficient framework, \holcon (Halo Optimised Lightcone Constructor), for post-facto construction of dark matter halo lightcones from simulations and use them to generate a mock galaxy catalogue. \holcon includes a module to optimise the lightcone’s orientation within the simulation box, minimising repeated structures when the survey volume exceeds a single box -- a common challenge in modern surveys.  A linear interpolation scheme tracks the evolution of halo properties across snapshots. Applied to the publicly available Uchuu simulation, we construct a lightcone of 50 ${\rm deg}^2$ and extending up to $z = 10$, providing representative coverage of deep fields of Stage IV surveys. We validate the lightcone for cosmological applications by comparing the dark matter halo clustering in the lightcone with those from the original simulation snapshots. Subsequently, we make the galaxy-halo connection on the lightcone with a redshift extended version of the {\scipic} algorithm producing a comprehensive set of descriptive galaxy attributes. \holcon leverages \dask, a scalable parallel computing pythonic framework for fast construction of dark matter halo lightcones enabling rapid creation of multiple statistical realizations  essential for robust cosmological inference. The produced galaxy mock makes predictions for clustering of \Halpha emitters, making it a useful cosmology resource.}   


\usepackage{natbib}
\bibpunct{(}{)}{;}{a}{}{,} 
\begin{document}
\maketitle
  \nolinenumbers

\section{Introduction}The study of the large-scale structure of the Universe, from the formation of the cosmic web to the evolution of galaxies, relies crucially on cosmological simulations. These simulations also provide the framework necessary to interpret the vast amounts of data collected by modern Stage IV astronomical surveys such as \Euclid  \citep{2011arXiv1110.3193L,EuclidSkyOverview}, DESI \citep{2016arXiv161100036D}, Rubin-LSST \citep{2019ApJ...873..111I}, the \textit{Nancy Grace Roman Space Telescope } \citep{2019arXiv190205569A}, and the China Space Station Telescope (CSST) \citep{2019ApJ...883..203G} which aim to map the Universe across extensive cosmological volumes and high density of sources, and to shed light on the nature of dark energy. Critical to this interpretation are mock catalogues, which serve as synthetic representations of the observed Universe, allowing testing and validation of analysis pipelines and constraining the errors on cosmological parameters with high precision. They facilitate the crucial pre-mission testing of data analysis pipelines and the assessment of the reliability of cosmological parameter forecasts. More recently, simulation-based inference approaches seek to directly infer cosmological parameters by leveraging large datasets of simulations to construct inference models \citep{2018MNRAS.476L..60A,2025arXiv250616408D}.

Since the primary motivation for using cosmological simulations is to extract realistic observables that can be directly compared with astronomical surveys, simulated mock catalogues require the construction of galaxies on a lightcone.
 There are fundamentally two approaches to constructing these lightcones. One method involves generating them on the fly during simulation run \citep{2002ApJ...573....7E,2008MNRAS.391..435F,2010MNRAS.401..705P,2015MNRAS.448.2987F,2021MNRAS.506.2871S}, which is computationally efficient and accurate. The alternative method is to perform an interpolation after the simulations have been completed \citep{2013MNRAS.429..556M,2019arXiv190608355H,2019A&A...631A..82I,2021JCAP...02..047S,2023MNRAS.525.4367H}. This post-processing approach provides significant advantages, including the flexibility to define custom orientations, specify survey geometries, and even select the observer’s position as the origin of the lightcone well after the initial simulation has been completed. However, this flexibility comes at a cost. It necessitates the construction of robust merger trees that accurately track haloes and galaxies across multiple simulation snapshots. This cross-snapshot tracking can be a numerically intensive search problem. In this work, we demonstrate that leveraging distributed computing systems \citep{10.5555/2228298.2228301,matthew_rocklin-proc-scipy-2015} provides an ideal solution to the computational challenges associated with matching objects across snapshots. This approach significantly expedites the lightcone generation process from pre-run simulations, making it possible to rapidly produce multiple realizations of lightcones, a crucial capability for robust statistical analyses and comparison with observational data.

Modern cosmological surveys, characterized by their extensive depth and wide angular coverage, exceed the typical dimensions of high-resolution cosmological N-body simulation boxes. State-of-the-art simulations require simultaneous resolution of small-scale structures (e.g., galaxies, dark matter haloes) and the inclusion of large cosmic volumes, which imposes a significant computational strain and challenge.

To generate mock observations that cover the extensive footprint of a survey, periodic simulation boxes are often repeated. This approach represents a fundamental problem in the sense that the scales larger than the box size are not present in the power spectrum or two-point correlation function. In addition to this, there are secondary problems, such as the repetition along certain directions that creates a visual "kaleidoscope effect" \citep{2005MNRAS.360..159B} and artificially suppresses the true cosmic variance for one-point statistics, such as the mass function, since the repeated structures do not represent statistically independent samples along the line of sight. Attempts to mitigate these artifacts include rotating the simulation boxes before repetition or stacking multiple realisations of the boxes side by side. However, this procedure introduces discontinuities in large-scale structures and affects two-point correlations at the box interfaces \citep{2014PhRvD..90b3520P} and also creates spurious cross-correlations across different redshift bins, although this can be practically tackled \citep{2025arXiv250712116E}.
\citet{2007MNRAS.376....2K} suggested orienting the lightcone towards the corners of the farthest repeated cube to minimize repetitive structures.
\citet{2009A&A...499...31H} and \citet{2010ApJS..190..311C} provided a mathematically robust methods for identifying skewed and elongated non-repetitive volumes, informed by the principles of crystal lattice theory. In this paper, we provide an alternative solution to address some of the problems of repeating boxes when constructing lightcone mocks.

Beyond the sheer scale of the surveys, another significant challenge in cosmological simulations for large-scale surveys is accurately modeling galaxy formation and baryonic physics. Partially or fully resolving the complex interplay of gas, stars, and feedback processes within a cosmological volume is computationally prohibitive, although significant progress in this direction has been achieved \citep{2023ApJS..265...54V,2023MNRAS.526.4978S}. Consequently, many simulations primarily run collisionless dark matter, and then employ `painting' techniques or semi-analytic models to connect galaxies to the dark matter haloes in the simulation. These techniques provide a computationally efficient way to populate dark matter haloes with realistic galaxy properties.

There are various techniques ranging from abundance matching (AM), sub-halo abundance matching  and halo occupation distribution (HOD) models \citep{2002PhR...372....1C,2004ApJ...609...35K,2013MNRAS.433..659H,2019MNRAS.488.3143B,2020MNRAS.499.4905C,2021MNRAS.503.4147P} to connect galaxy properties with halo properties which are employed in survey mocks \citep{2023MNRAS.519.1648A,2024A&A...691A.136L,2025arXiv250712116E}. In this context, the Scientific Pipeline at PIC \citep[{\scipic},][]{2017ehep.confE.488C} is a powerful suite of algorithms integrated into a pipeline for generating extensive synthetic galaxy catalogues, providing over 200 descriptive properties for each galaxy. It provides a galaxy-halo connection using a simple HOD and employs abundance matching (AM) to assign luminosities to galaxies. {\scipic} is utilized in the dark matter only simulation called the \Euclid flagship \citep{2017ComAC...4....2P} to generate synthetic galaxy mocks, especially for lensing and clustering studies \citep{2025A&A...697A...5E}. 

As modern deep surveys such as CANDELS \citep{2011ApJS..197...35G,2011ApJS..197...36K,2017ApJS..228....7N}, COSMOS \citep{2007ApJS..172....1S,2016ApJS..224...24L,2022ApJS..258...11W} and JADES \citep{2023ApJS..269...16R} continue to probe deeper, thus reaching higher redshifts, our ability to accurately study galaxy formation and evolution relies on efficient and robust data analysis pipelines. To validate and optimise such pipelines for future deep survey observations, a realistic mock catalogue is indispensable. Here, we will also describe the extension of the {\scipic} pipeline with the aim of designing a mock catalogue that replicates the properties of a deep galaxy survey. We will then apply the pipeline to a dark matter lightcone constructed having a sky coverage of around $50\, \rm{deg}^2$ to be representative of stage IV cosmological surveys deep field observations like in \Euclid \citep{2025A&A...695A.259E,EuclidSkyOverview}, Rubin-LSST \citep{2022ApJS..258....1B,2024ApJS..275...21G} and ODIN \citep{2024ApJ...962...36L}. 

In a realistic flux-limited survey, the observed galaxy samples to be studied represent biased tracers of the underlying dark matter distribution. Emission-line galaxies are an efficient tracer of the matter distribution since their redshifts can be accurately determined, and the deduced spatial distribution can be used to constrain cosmological models. \Euclid and the \textit{Nancy Grace Roman Space Telescope} will carry out slitless spectroscopy in the near-infrared to identify emission-line galaxies, with a particular emphasis on \Halpha emitters. Spectroscopy in the near-infrared enables coverage of a redshift range that is challenging to access from ground-based facilities, thereby providing cosmological constraints during an epoch that remains largely unexplored. Approaches to model the nature and clustering of the emission line galaxies range in complexity from simulations including complex physics of radiative transfer, semi-analytic approaches that partially calibrate parameters with observations to HOD based approaches \citep{knebe2020,Nusser_2020,tacchella,2024A&A...689A..66O,2024MNRAS.529.3877R,2025ApJ...988...44R}.  In this work, we will simulate the selection of a flux-limted sample of \Halpha emitters as an indication of what can be achieved with the new near-infrared spectroscopic surveys from space.
 
This paper is organized as follows. \Cref{sec:code} introduces the method used to construct halo lightcones. \Cref{sec:ooa} details the optimization of the lightcone orientation for a given geometrical requirement of sky coverage and redshift depth, to maximally probe unique structures along the central line-of-sight direction of the lightcone. \Cref{sec:lcgen} discusses the interpolation scheme and distributed computing setup to facilitate tracking haloes across snapshots.
In \cref{sec:lcuchuu}, we use \holcon to construct a single realisation of the lightcone using the publicly available Uchuu simulations and perform validation tests to assess its suitability for clustering studies. In \cref{sec:gal}, we demonstrate the utility of the constructed lightcone by populating it with galaxies by running a modified version of {\scipic} extended to $z = 10$ and present the clustering predictions for a sample of \Halpha emission-line selected galaxies. \Cref{sec:sumamry} provides a summary and discusses future directions. 
We have made our code for building a lightcone publicly available.\footnote{\url{https://github.com/rsujatha/HOLCon}}

\section{Halo lightcone construction}
\label{sec:code}
In this section, we introduce \holcon (Halo Optimised Lightcone Constructor), a tool designed to construct lightcones from cosmological simulation snapshots. \holcon also includes a utility code to determine optimal orientations of the lightcone in order to maximize survey volume and minimize duplication of structures, and provides a systematic method to generate post facto lightcones from existing snapshot data. In the first subsection, we derive the analytic mask function used to identify orientations with minimal repetition, followed by a detailed description of the procedure for constructing the halo lightcone in the subsequent subsection. 
\subsection{Optimal orientation angles}
\label{sec:ooa}

\begin{figure}[hbt!]
\centering
\includegraphics[width=0.95\linewidth]{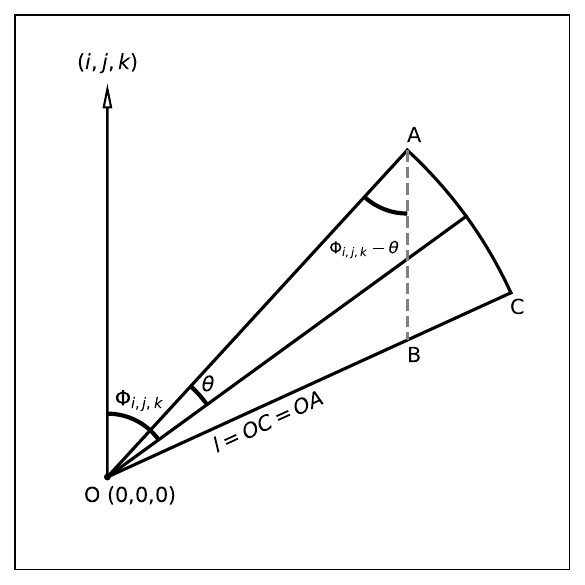}
\caption{Diagram of the lightcone placed with the observer at the origin. It is oriented at an angle with the direction vector (${i},{j},{k}$) where $i,j,k$ are integers and point to one of the vertices of the primary simulation cube and subsequent repeated cubes. Here, $\theta$ is the angular radius and $l$ is the length of the lightcone. The diagram shows the case where the lightcone is oriented at a limiting angle $\Phi_{i,j,k}$ with respect to the direction $(i,j,k)$. Any angle less than the limiting angle results in repetition of structures and an angle larger is free of repeated structures associated with this direction.}
\label{fig_sym}
\end{figure}

In this subsection, our goal is to identify the optimal angle to embed a lightcone inside a simulation box in order to minimize/eliminate the repetition of cosmic structures.
We will consider a simple geometry for the lightcone with a circular footprint characterized by the angular radius $\theta$. The maximum redshift probed by the lightcone corresponds to a comoving length, $l$. A cross-sectional view of the lightcone is shown in \cref{fig_sym}, where the observer is placed to coincide with the origin or vertex of the simulation cube.

When the comoving length, $l$, is shorter than the simulation box-size $L$, we are guaranteed to span unique structures inside the lightcone. However, for deeper lightcone geometries ($l> L$), we are faced with having to repeat the periodic boxes multiple times. At the boundaries of periodic boxes, the large-scale structures and the cosmic-web seamlessly interconnect, preserving the n-point correlations at scales less than the box length $L$. 

\begin{figure}
\centering
\includegraphics[width=0.99\linewidth]{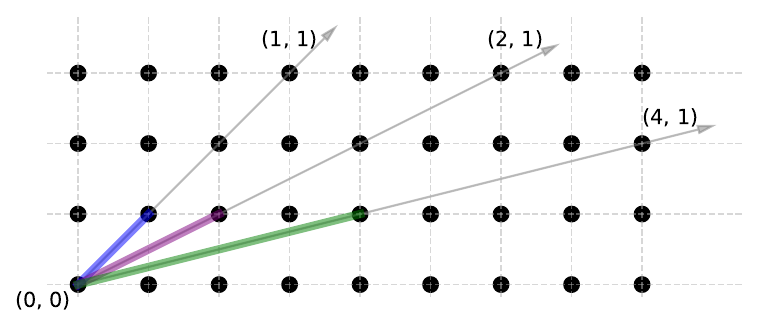}
\caption{Two-dimensional lattice points representing the corners of the periodic simulation box and the subsequent repeated boxes placed side-by-side. The maximum length that probes unique structures along a direction or `{the repetition length}' is shown in blue, purple and green colors corresponding to the directions (1,1), (2,1), (4,1) respectively.}
\label{fig_lattice}
\end{figure}

Consider repeated periodic boxes to be part of a lattice structure (see \cref{fig_lattice} for a two-dimensional representation). The corners of the original box and the subsequent boxes are denoted by $(i,j,k)$ where $i,j,k$ are integers starting from $i=j=k=0$.
It is intuitive to see that a pencil beam lightcone starting at the origin/vertex of the simulation when directed towards the other points of the cubic lattice will face repetition in structures only after crossing the nearest lattice point in that direction. This allows us to associate a `repetition length', $L^2(i^2+j^2+k^2)$ to every direction $(i,j,k)$,\footnote{To avoid counting the same directions twice, here we assume that $i,j,k$ do not have common multiples.} which is the maximum length for one-dimensional unique structures along that direction after which the exact structures will repeat \citep[see]{2009A&A...499...31H}. \Cref{fig_lattice} shows three such lattice directions and the repetition length associated with those directions in blue, purple and green lines. To identify problematic directions, we only need to consider the nearby lattice points, i.e., those whose repetition length is less than the length of the lightcone $l$, which translates to the condition $(i^2+j^2+k^2)<l^2/L^2$. A numerical search for the orientation directions of zero repetitions has been done in \cref{app:numerical}. This search is  consistent with the results of \cite{2024MNRAS.534.1205C}, and it shows that indeed the angles where repetition occurs are zones around the direction of the lattice points $(i,j,k)$, with the angular radius $\Phi_{i,j,k}$ parametrised by the geometry of the lightcone (See \cref{fig:numerical}). 

This motivates us to define a mask function $M$ that identifies regions where constructing a lightcone would result in repeated structures and assigns them a non-zero value, 
\begin{equation}
M := \sum_{i,j,k}^{(i^2+j^2+k^2)<l^2/L^2} H\left(\Phi - \Phi_{i,j,k}\right)\,,
\label{eq:sketch}
\end{equation}
where $\Phi$ is the angle made by the lightcone with respect to the $(i,j,k)$ direction, $H$ is the Heaviside step function that takes a value equal to one for $\Phi > \Phi_{i,j,k}$ and a value 0 for $\Phi < \Phi_{i,j,k}$.
For convenience we would like to express the mask function in terms of the right ascension, $\alpha$, and the declination, $\delta$, which defines the direction the center of the lightcone relative to the simulation box axes. We can express $\Phi$ with respect to these angles as,
\begin{equation}
\Phi = \arccos{\left(i\cos{\alpha}\cos{\delta} + j \sin{\alpha}\cos{\delta} + k\sin{\delta}\right)}\,.
\label{eq:phi}
\end{equation}

\begin{figure*}
  \begin{subfigure}[b]{1.0\textwidth}
\centering
\fbox{\includegraphics[width=0.9\textwidth]{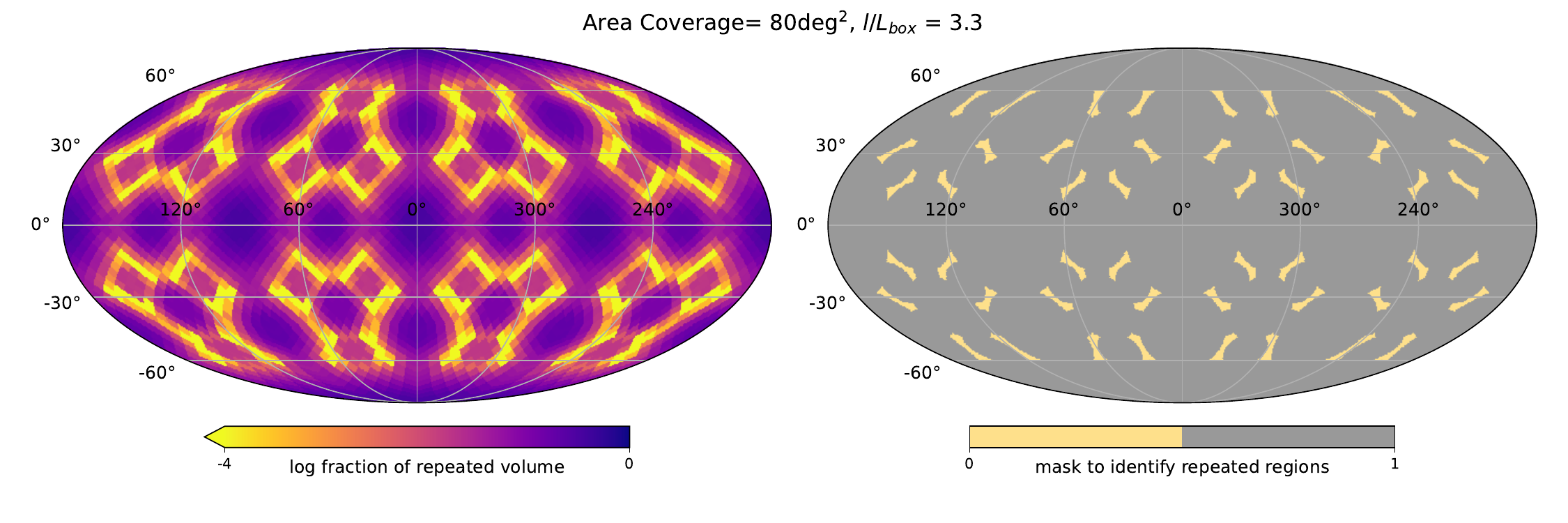}}
  \end{subfigure}
\begin{subfigure}[b]{1.0\textwidth}
\centering
\fbox{\includegraphics[width=0.9\textwidth]{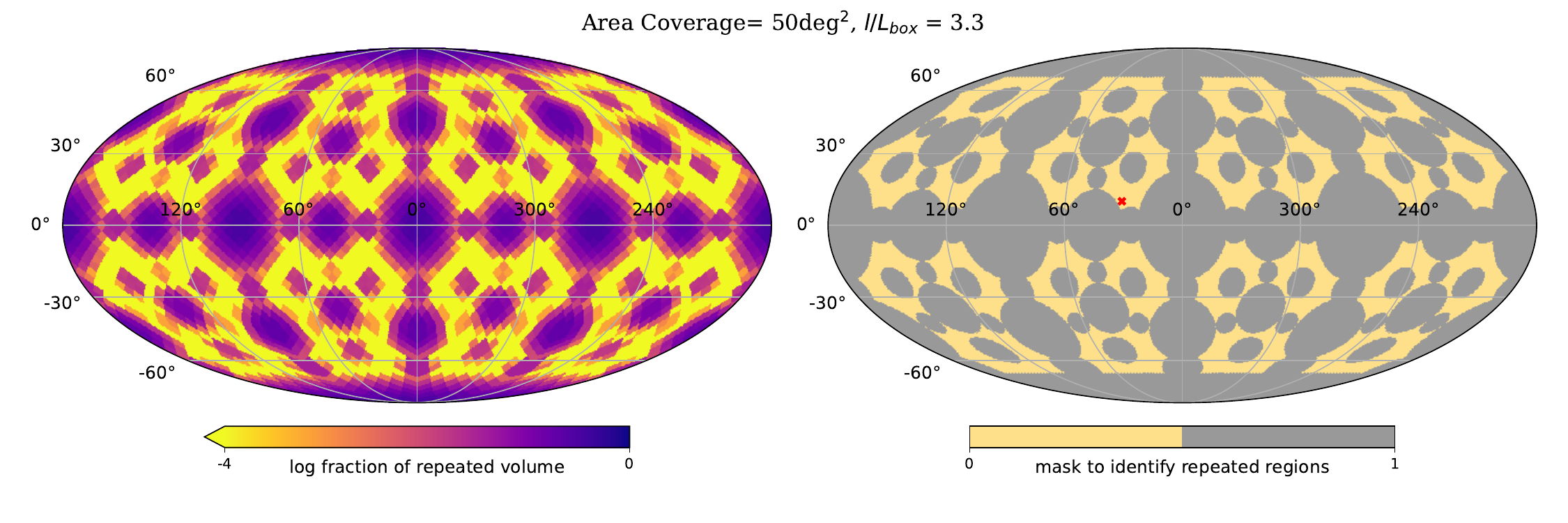}}
\end{subfigure}
\begin{subfigure}[b]{1.0\textwidth}
\centering
\fbox{\includegraphics[width=0.9\textwidth]{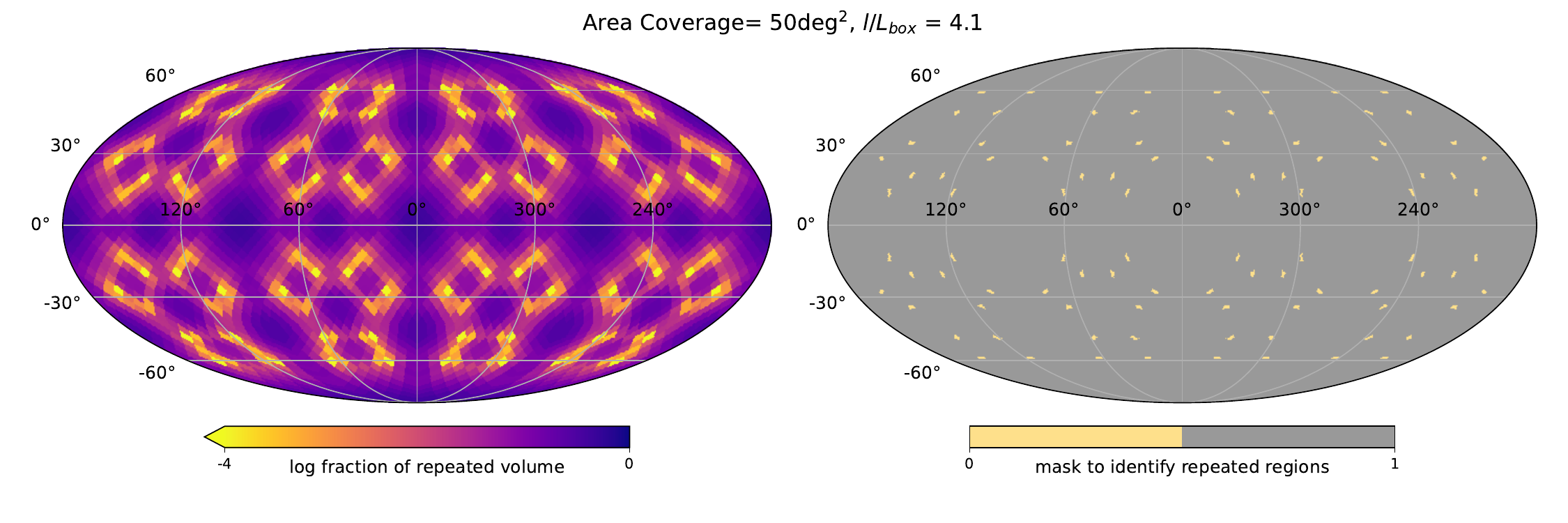}}
\end{subfigure}

\caption{Mollweide projection of the sky regions of repetition inside the periodic box: the left panel shows the numerically computed fraction of volume that is repeated and the right panel shows the analytic mask that demarcates lightcone orientations with repeated volumes from unique volumes. The top, middle and bottom panel corresponds to three different configurations of area coverage and $l/L_{\rm box}$. We can see that the region with unique structures (yellow) progressively decreases as we increase the sky coverage or increase $l/L_{\rm box}$. We have used \cref{eq:main} to plot the right panels. The red cross mark in the middle right panel denotes the chosen orientation for constructing a lightcone later in the text (See \cref{sec:lcuchuu}). An interactive version of the mask function is available at \url{https://rsujatha.github.io/HOLCon}.} 
\label{fig:config1}
\end{figure*}

 In the following, we will compute the exact value for $\Phi_{i,j,k}$.
For demonstration, the lightcone is precisely oriented at the limiting angle $\Phi_{i,j,k}$ with respect to $(i,j,k)$ in \cref{fig_sym}. At this limiting angle, the largest length along $(\hat{i},\hat{j},\hat{k})$ that can be accommodated without structural repetition inside the lightcone ($AB$) is equal to the repetition length associated with this direction, i.e.,
\begin{equation}
  AB^2 = L^2(i^2+j^2+k^2)\,,
  \label{eq:condition}
\end{equation}
Using the law of the sines on the triangle $\triangle ABC$ combined with the condition in ~\cref{eq:condition}, $\Phi_{i,j,k}$ can be expressed in terms of the angular radius of the lightcone $\theta$, and its relative length in relation to the box size $l/L$,
\begin{equation}
  \Phi_{i,j,k} = \arcsin\left( \dfrac{(l/L)\sin{2\theta}}{\sqrt{i^2+j^2+k^2}}\right)-\theta \,.
  \label{eq:condition1}
\end{equation}
It is straightforward to determine the mask function in \cref{eq:sketch} from \cref{eq:phi} and \cref{eq:condition1}. The mask function can also be expressed as the difference in the cosines of the two angles $\Phi_{i,j,k}$ and $\Phi$, in which case it becomes,

\begin{equation}
\begin{aligned}
  M(\alpha,\delta) = & \sum_{i,j,k}^{(i^2+j^2+k^2)<\frac{l^2}{L^2}} H \bigg( 
  k\sin{\delta}+\cos{\delta}\,(i\cos{\alpha}+j \sin{\alpha}) \\
  &-(l/L)\sin{2\theta}\sin{\theta} \\
  &-\cos{\theta}\sqrt{i^2+j^2+k^2-(l/L)^2\sin^2{2\theta}} 
  \bigg)\,.
\end{aligned}
  \label{eq:main}
\end{equation}
In the right panels of \cref{fig:config1} we show various configurations of the mask function, i.e., we assign different values for $\theta$ and $l/L$ and plot the resulting $M(\alpha,\delta)$. The right panels show the mask calculated using the analytical expression from \cref{eq:main}. The left panels show a numerical calculation of the fraction of volume of repeated structures with the same lightcone configurations as in the right panels. We can see that both methods produce equivalent mask functions. The numerical approach in the left panels was introduced earlier in \citet{2024MNRAS.534.1205C} and has been used in the generation of CSST mock samples and emulators \citep{2025SCPMA..6889512C,2025SCPMA..6809511H,2025ApJ...985..131X}. In this section, we have presented an analytical version of the same computation that it is not only consistent with the numerical code of \citet{2024MNRAS.534.1205C} but also provides a closed-form analytic solution that expedites the production of mock catalogues.


\subsection{Construction of haloes in the lightcone}
\label{sec:lcgen}
This section details the methodology used to construct the lightcone halo catalogue from a series of simulation snapshots. We begin by describing the interpolation scheme used to place haloes that cross the lightcone at their correct positions. Following that, we discuss strategies for defining a buffer zone to account for haloes moving in and out of the sky coverage volume. Finally, we explain how we use distributed computing frameworks like Apache Spark and \dask to overcome memory limitations arising from handling massive simulations and efficiently perform cross snapshot connection between haloes required for interpolation.

\subsubsection{Interpolation scheme}

We will first describe the interpolation scheme used to describe the halo in between snapshots.
The halo properties are linearly interpolated between the two snapshots using a scheme similar to the one described in \citet{2022MNRAS.509.2194H}. Although higher order interpolation schemes such as cubic have been explored previously \citep{2013MNRAS.429..556M}, it was also shown that such interpolations produce extreme velocities between snapshots and using linear interpolation improves the correlation function quadrupole measurements on small scales \citep{2022MNRAS.516.4529S}. In the following, we will use subscripts {\rm i} and f to denote quantities in the initial and final snapshots between which the haloes are linearly interpolated. When the time interval between the snapshot is minimal, we can assume that the haloes, at first order, move linearly with constant velocity $V$,
\begin{equation}
V = (r_{\rm f}-r_{\rm i})/(\chi_{\rm f}-\chi_{\rm i}),
\end{equation}
where $r_{\rm f}$ and $r_{\rm i}$ are the position of the same halo in the two snapshots and $\chi_{\rm f}$, $\chi_{\rm i}$ are the comoving distances used as proxies for time at the final and initial snapshots. The halo crosses the observer's lightcone somewhere in between $r_{\rm f}$ and $r_{\rm i}$, at some time corresponding to $\chi_{*}$, and its position is $r_{*}=\chi_{*}$. The velocity can also be written in terms of $\chi_{*}$
\begin{equation}
V = (r_{\rm f}-\chi_{*})/(\chi_{\rm f}-\chi_{*}),
\end{equation}
From the above two equations, the position where the halo crosses the lightcone is computed to be
\begin{equation}
    \chi_{*} = (r_{\rm f} - V \chi_{\rm f})/ (1-V),
    \label{eq:halocross}
\end{equation}
or equivalently,
\begin{equation}
    r_{*} = r_{\rm f} + \dfrac{\chi_{\rm f}-\chi_{*}}{\chi_{\rm f}-\chi_{\rm i}}(r_{\rm i}-r_{\rm f})\,,
    \label{eq:posintpol}
\end{equation}
Let $q_{\rm i}$ and $q_{\rm f}$ be the corresponding values of some generic halo property at the initial and final snapshot respectively; we adopt the following interpolation, which is the same as that used for interpolating halo positions in \cref{eq:posintpol},
\begin{equation}
    q_{*} = q_{\rm f} + \dfrac{\chi_{\rm f}-\chi_{*}}{\chi_{\rm f}-\chi_{\rm i}}(q_{i}-q_{\rm f})\,,
\end{equation}
where $q_{*}$ is the value of the halo property when the halo crosses the lightcone to the observer at a comoving distance of $\chi_{*}$ given by \cref{eq:halocross}.\\

 In order to achieve this interpolation, haloes across snapshots are matched with the progenitors and descendants. For example, in \texttt{ROCKSTAR} \citep{2013ApJ...762..109B}, which serves as our default halo finder, the merger trees are identified with \texttt{CONSISTENT TREES} \citep{2013ApJ...763...18B} that track the haloes across snapshots. Each halo in a snapshot is provided with a  {\texttt{desc\_id}} that maps to the {\texttt{id}} of the descendant halo in the future snapshot. Similarly, mapping the {\texttt{last\_mainleaf\_depthfirst\_id}} of a halo to the {\texttt{last\_mainleaf\_depthfirst\_id}} of the halo from the past snapshot ensures that haloes are matched to the most massive progenitor. (See, for example, \cite{2016MNRAS.462..893R} for a description of the different merger tree identifiers.)
We also integrate support for the \textsc{hbt-herons} halo finder and merger tree framework \citep{2025arXiv250206932F} where the \texttt{TrackId} is used to track the subhaloes across snapshots. In this case, haloes are tracked across snapshots using the \texttt{TrackId} field. When a subhalo undergoes a merger, its \texttt{TrackId} automatically maps to the \texttt{TrackId} of its most massive progenitor in the previous snapshot, while the \texttt{DescendantTrackId} links it to the descendant in the next snapshot.  Since the exact timing of mergers within the interval between snapshots is unknown, we assume that mergers occur halfway in comoving distance between the snapshots. This assumption can introduce a jump discontinuity in redshift at the high-mass end of the halo mass function, which may be mitigated by using more sophisticated interpolation techniques.

Some haloes are born in the final snapshot and are entirely absent in the previous snapshot. These can be identified by \texttt{num\_prog == 0} in the \texttt{CONSISTENT TREES} catalogue and using the \texttt{SnapshotIndexOfBirth} in the case of \texttt{HBT-HERONS}. In principle, these haloes could have formed at any time before the final snapshot, up to the epoch of the previous snapshot. To model this, we assign a probability of birth that decreases as the formation time moves further into the past, relative to the redshift of the snapshot. Using this probability, we populate the lightcone with newly formed haloes. This approach ensures a smooth transition at the low-mass end of the halo mass function in the lightcone.

\subsubsection{Edge case handling}

\label{subsec:edge}
For relatively small footprints, it is also important to define a buffer size around the area of sky coverage while running our constructor. To identify relevant haloes for matching, our algorithm queries regions in the initial and final snapshots that are likely to intersect the lightcone. Then a one-to-one matching is performed between the haloes in these snapshots. However, some haloes may move out of or into these regions between snapshots. To account for such cases and ensure accurate matching and interpolation, we define a buffered volume extending beyond the initially identified regions. The buffer size is a rough estimate based on the maximum possible displacement of haloes between snapshots. Since we are interested only in parent/main haloes that host central galaxies, we consider the cosmic velocity dispersion, which is taken approximately to be $\sigma_{\rm v} = 350 \kms$, as a good number to estimate the buffer size.\footnote{The subhaloes that host satellite galaxies can have complex non-linear motion between snapshots. They can revolve around the potential wells of the centrals. In our galaxy populating pipeline described in \cref{sec:gal}, we place satellite galaxies following a theoretical profile (See Section 5.3 of \citealt{2025A&A...697A...5E}). Therefore, we do not need to
track satellites/subhaloes and we do not consider their velocities here for estimating buffer size.} We also use the maximum cosmic time interval between adjacent snapshots $\Delta {T}$ to estimate the buffer size as this would give a conservative upper limit. The buffer size is then calculated as the product of the time interval between snapshots and the typical velocity dispersion of the haloes i.e., ${\rm BufferSize} = \Delta {T} \times  \sigma_{\rm v}$. (See \cref{sec:lcuchuu} for the actual numbers that we use in this work.)
\subsubsection{Distributed halo matching and interpolation: Beyond
in-memory constraints}
With unprecedented volumes and resolutions, modern cosmological simulations (e.g., Flamingo, UNIT, Euclid-Flagship, Uchuu) generate individual halo catalogues that reach terabytes per snapshot, surpassing the memory limits of normal computers. The task of robustly and efficiently identifying descendant haloes from one snapshot to the next presents computational challenges. Manually distributing the halo matching logic across multiple compute nodes, handling data partitioning, communication, and fault tolerance, is a non-trivial engineering task. 
To move beyond in-memory constraints and unlock the scalability required for these analyses, we utilize high-level distributed computing frameworks which are purposefully built to handle datasets that exceed the memory of a single machine, automatically managing data partitioning, distributed execution, and efficient out-of-core processing when necessary. 

Here, we use two main distributed computing frameworks: Apache Spark and \dask.
Apache Spark is an open-source system for large-scale data processing. Its main concept, the Resilient Distributed Dataset, is an immutable, fault-tolerant collection of data that can be processed in parallel. Spark also includes higher-level tools such as Spark SQL for structured data, Spark Streaming for real-time processing, and MLlib for machine learning, making it a complete platform for data-intensive tasks.
\dask is an open-source Python library for parallel computing. Unlike Spark, which is a standalone system, \dask is designed to scale existing Python tools like NumPy, Pandas, and Scikit-learn. It represents computations as task graphs, which can run on a cluster or a single multicore machine. Because it’s lightweight and Python-centric, \dask is especially suitable among researchers and data scientists who want to scale their Python workflows easily.

 The distributed computing frameworks are particularly useful in enabling us to perform complex relational operations like distributed \texttt{JOIN} on these massive datasets and connect haloes to their progenitors in the past snapshot and their descendants in the future snapshot. \Cref{tab:joins} provides details of the necessary \texttt{JOIN} operations between various halo merger trees IDs  belonging to adjacent snapshots in order to interpolate between them. 
In this work, we have generated a $50\,{\rm deg}^2$ lightcone using a \dask distributed computing framework deployed through HTCondor. Using 50 workers, the lightcone is constructed in less than 10 minutes, demonstrating the feasibility of generating multiple statistical realizations.
%%% here DASK is in capitals before it was Dask. Make consistent all references 
%%% 10 minutes does not mean much by itself. How much CPU time in which type of machine/processor?

\begin{table*}[htbp]
    \centering
    \caption{Summary of the \texttt{LEFT JOIN} Operations with Halo Matching and Interpolation.}
    \label{tab:join_operations}
    \begin{tabular}{llcc}
        \toprule
        \textbf{Halo Finder} & \textbf{Halo relationship} & \textbf{Left Table ID} & \textbf{Right Table ID} \\
        \midrule
        \addlinespace[0.5em] % Add some vertical space
        \textbf{\texttt{ROCKSTAR}}\\+\textbf{\texttt{CONSISTENT TREES}} & Descendants (N to N+1) & \texttt{\texttt{desc\_id}} & \texttt{id} \\
                         (Uchuu Simulation) & Progenitors (N+1 to N) & \texttt{last\_mainleaf\_depthfirst\_id} & \texttt{last\_mainleaf\_depthfirst\_id} \\
        \addlinespace[0.5em] % Add some vertical space
        \textbf{\textsc{HBT-HERONS}}    & Descendants (N to N+1) & \texttt{TrackId} & \texttt{TrackId} \\
                       (Flamingo Simulation)   & Progenitors (N+1 to N) & \texttt{TrackId} & \texttt{TrackId} \\
        \bottomrule
    \end{tabular}
\label{tab:joins}
\end{table*}
\section{Data and Validation}
\label{sec:lcuchuu}
\subsection{Simulations - Uchuu halo lightcone}
In this section, we apply the method described above to generate a lightcone for the Uchuu simulation.  The Uchuu suite comprises a set of ultra-large, high‑resolution cosmological $N$‑body simulations based on the Planck 2015 cosmology \citep[$\Omm=0.3089, \Omega_{\Lambda}=0.6911, h=0.6774, \sigma_{8}=0.8159, n_{s}=0.9667$,][]{2016A&A...594A..13P}. The flagship run, which we utilize here, evolves 2.1 trillion $(12800^3)$ dark‑matter particles in a comoving cube of side $L = 2.0\, h^{-1}\,\mathrm{Gpc}$, yielding a particle mass of $m_{\rm p} = 3.27\times10^8\,h^{-1}M_\odot$. Publicly available catalogues are hosted on the Skies and Universes website,\footnote{\url{https://skiesanduniverses.org/Simulations/Uchuu/}} and also hosted on CosmoHub.\footnote{\url{https://cosmohub.pic.es/catalogs/316}}\citep{2020A&C....3200391T} 

\begin{figure}
    \centering
\includegraphics[width=0.87\linewidth,trim={14cm 10.4cm 15.5cm 7cm},clip]{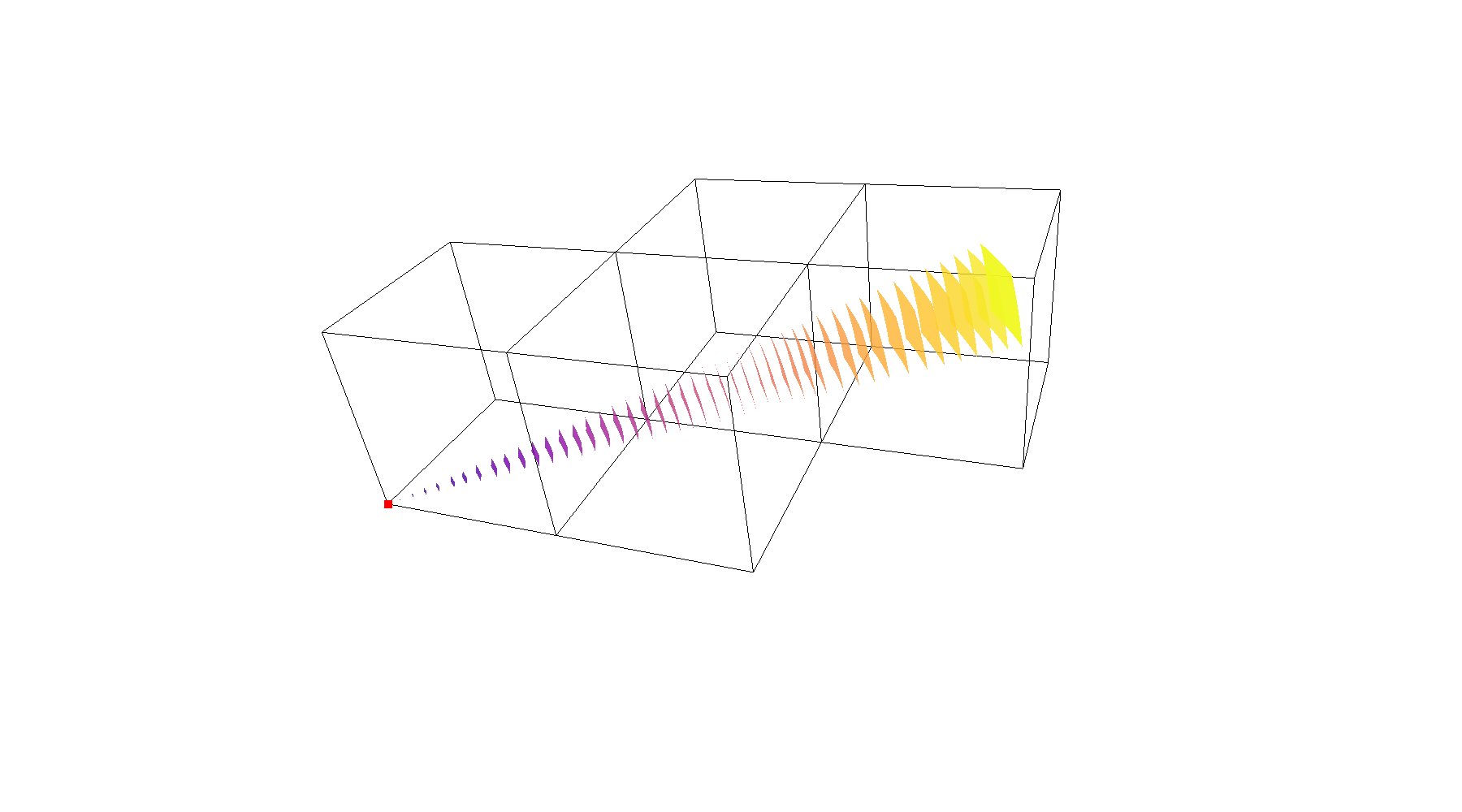}
    \includegraphics[width=0.12\linewidth]{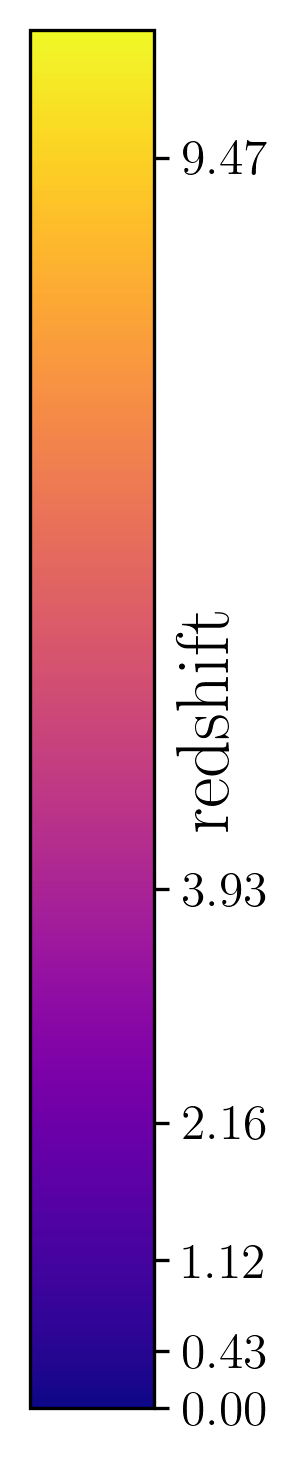}
    \caption{Schematic representation of the lightcone configuration: The sky coverage is 50 $\mathrm{deg}^2$ going up to depth of $z\sim 10$. The red cross in the middle panel of \cref{fig:config1}, marking a right ascension of $\alpha = 31 \degree$ and a declination of $\delta = 10  \degree$, indicates the chosen orientation for this lightcone with respect to a cartesian coordinate system oriented along the sides of the periodic box. Although 4 box repetitions are required to reach the highest redshift, the chosen orientation guarantees that the lightcone probes unique large-scale structures and haloes.}
    \label{fig:schematic}
\end{figure}

In this work, we used a subset of 47 snapshots and halo catalogues covering redshifts from $z=0$ to $z = 10$. Here we are interested in constructing a lightcone covering a circular footprint area of $50~\mathrm{deg}^2$ which is a typical size of deep fields in stage IV cosmological surveys. Since this is a relatively small but deep footprint, there is considerable freedom in how to orient the lightcone within the periodic simulation box. 
Based on the discussion in the previous section, we select an orientation that probes unique large-scale structures and haloes while minimizing the number of repeated boxes along the line of sight. For the cosmology of the simulation, a redshift of $z = 10$ corresponds to a comoving distance of $l = 6.5~h^{-1}~\mathrm{Gpc}$, giving a ratio of $l/L = 3.3$.  Substituting these parameters into \cref{eq:main} results in the mask function shown in the middle right panel of \cref{fig:config1}, where the yellow regions denote the optimal orientations. Given our goal to minimize box repetition while probing unique structures, this leads us to select an orientation with right ascension $\alpha=31^\circ$ and declination $\delta=10^\circ$, marked by a red cross in \cref{fig:config1}. A schematic representation of this orientation of the lightcone is shown in \cref{fig:schematic}, embedded inside 4 periodic boxes. The observer is denoted by a red dot and has been placed at the origin of the first periodic box. The patches, each representing a small section of a spherical shell, show where different stored snapshots intersect the observer’s lightcone. The redshift for each patch is given by the color bar on the right. 
To estimate the buffer size discussed in \cref{subsec:edge} the maximum time interval between adjacent snapshots is 0.3 Gyr. Multiplying this by the velocity dispersion of 350 \kms gives a characteristic distance scale of approximately $0.22\, \hMpc$. Although this is smaller than $1\, \hMpc$, for order-of-magnitude estimates and to maintain a conservative buffer, we approximate this as roughly $1\, \hMpc$. 

In \cref{app:validation} we have performed various tests to validate the use of the halo lightcone for cosmological purposes. In \cref{fig:massfn} we have computed the halo mass function (HMF) in 20 equally spaced bins in redshift between 0 and 10 in the lightcone and compared it against the mass function from the simulation snapshots. 
 We also compare the two-point correlation function of the dark matter haloes in the lightcone vs snapshots to validate its use for cosmological analysis.  We had also performed tests to compare the concentration-mass relation and other halo properties of the lightcone in narrow redshift bins against the Uchuu snapshots at the same redshift, but these are not shown for brevity.

\subsection{Observations - COSMOS2020}
To calibrate and validate the galaxy mocks below, we use COSMOS2020, a comprehensive multi-wavelength catalogue reaching out to high redshift \citep{2022ApJS..258...11W}. We use the Farmer version with \texttt{COMBINED\_FLAG=0} and a sky coverage of 1.27 $\rm deg^2$. The catalogue is also processed to correct for the Milky Way attenuation. Photometric redshifts estimated from fitting spectral energy distribution (SED) templates with \texttt{Lephare} \citep{2011ascl.soft08009A} are used as an estimate of the true redshift.
\section{Application: {\scipic} galaxy catalog}
\label{sec:gal}
In this section, we populate the Uchuu lightcone with galaxies using a pipeline based on \scipic ~that provides synthetic galaxy catalogues taking into account a set of halo properties. First developed in \citet{2017ehep.confE.488C} for the MICE Grand Challenge simulations \citep{2015MNRAS.448.2987F,2015MNRAS.453.1513C,2015MNRAS.447.1319F} and extended in \citet{2025A&A...697A...5E} for use in the Euclid Flagship simulation \citep{2017ComAC...4....2P}, the pipeline aims to provide a comprehensive and realistic mocks that mimic the observed number density and galaxy clustering for flux limiting surveys. In order to do that, it makes the galaxy-halo connection using a simple HOD prescription. Each halo is always assigned one central galaxy above the mass threshold corresponding to the minimum halo mass in the simulation, i.e., $6.54\times 10^{8} M_{\odot}h^{-1}$. 
%%% I would delete with certainty. You can say "Each halo is always assigned one central galaxy" instead.
The number of satellites is Poisson distributed such that the average number of satellites hosted by a halo of mass $M_{\rm h}$ is given by 
\begin{equation}
    \langle{N_{\rm sats}\rangle} = \left(\dfrac{M_{\rm h}}{f M_{\rm min}}\right)^{\alpha_{\rm HOD}}\,,
    \label{eq:satellite}
\end{equation}
where $M_{\rm min}$ is the minimum mass of the simulation and $\alpha_{\rm HOD}$ and $f$ are parameters constrained to match the observed luminosity functions and the clustering at low redshift. Redshift evolution in the mocks is effectively encoded through the evolution of the HMF and through the observed luminosity functions over the redshift range considered. This allows making predictions at higher redshifts.
Further down the pipeline, each galaxy is assigned an \textsc{sed} that allow us to compute a set of fluxes/magnitudes in different filters. The mock observations are calibrated to reproduce basic properties like the overall galaxy number counts and color distributions \citep[COSMOS2020,][]{2022ApJS..258...11W}, as well as specific properties such as the number counts and clustering of certain tracers like \Halpha emitters. We will hereafter refer to this as the default pipeline. For a complete documentation of the default pipeline, we refer the reader to \cite{2025A&A...697A...5E}.
In this work, we have recalibrated the pipeline to apply it to the Uchuu simulations and to extend it to higher redshifts.
\subsection{Recalibration of the pipeline for the Uchuu Simulation}
\begin{figure}
    \centering
    \includegraphics[width=0.99\linewidth]{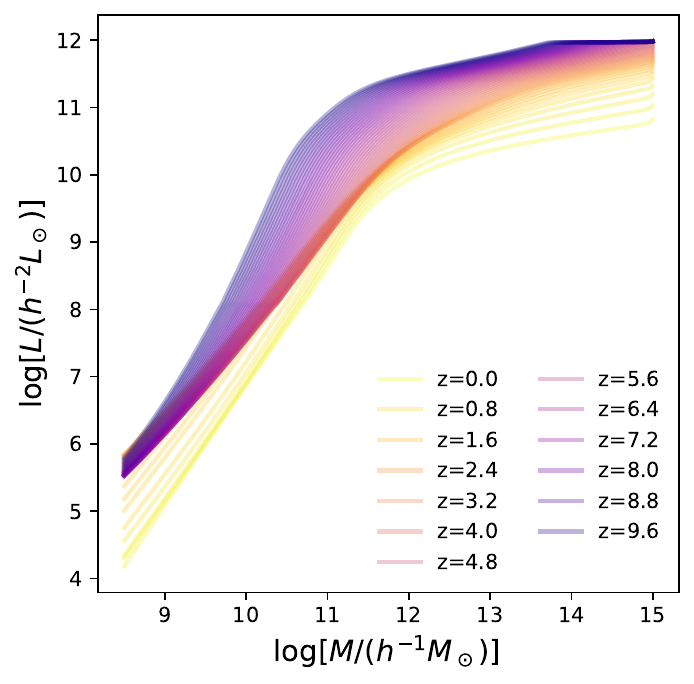}
    \caption{Relation between halo mass and luminosity at various redshifts
resulting from abundance matching the cumulative galaxy function and
the cumulative unscattered luminosity function.}
    \label{fig:mass-lum-relation}
\end{figure}

Here, we summarize the changes done to the default pipeline to adapt it to the Uchuu simulation. 
\begin{itemize}
\item We characterise the HMF from Uchuu for the entire halo mass and redshift range using a 5-parameter Schechter-like function (see \cref{app:validation} for details).
    \item At the HMF low-mass end, halo masses are discontinuous and incomplete due to the resolution limit of the simulation. 
    In order to correct for the incompleteness, we perform an AM transformation on the  original masses less than a threshold so that the final HMF after the transformation is complete. The discreteness of the halo mass function at very low masses (corresponding to haloes made up of only a few tens of particles) is corrected by uniformly redistributing halo masses between $M_{\rm h}$ and $M_{\rm h}+m_{\rm p}$ smoothing out the otherwise discrete spikes (see \cref{app:masscorrection} for details).

\item The halo mass function and the halo occupation description are combined to obtain a cumulative galaxy function. The $r01$-band\footnote{This is same as the $^{0.1}r$ notation in \citet{2003ApJ...592..819B}}  luminosities are assigned based on abundance matching the luminosity function with the cumulative galaxy function.
\Cref{fig:mass-lum-relation} shows the resultant relation monotonic relation between the halo mass and luminosity. While assigning galaxies a value for their $r01-$band luminosity, a Gaussian scatter of $\sigma_{L}$ is added in addition to the halo mass - luminosity relation.
\item The HOD parameters in \cref{eq:satellite} have been recalibrated from \( f = 15 \) to \( f = 22.4 \) and $\alpha_{\rm HOD}$ from 1.0 to 1.05. The Gaussian scatter $\sigma_L$ applied to the central luminosities is calibrated to the same value of 0.12 as in the default pipeline. Since the combination $fM_{\rm min}$ corresponds to the halo mass at which haloes host typically one satellite galaxy and since $M_{\rm min}$ is slightly larger for Uchuu, the increase in the value of $f$ reflects primarily the change in the simulation mass resolution (the particle mass is $m_{\rm p} = 3.27\times 10^{8} \,{\rm M_{\odot}}\, h^{-1}$ in Uchuu and $m_{p}=10^9\, {\rm M_{\odot}}\, h^{-1}$ in Flagship 2).\footnote{The $fM_{\rm min}$ is not the same in both the simulations. Since our mass correction procedure resets the minimum mass of the simulation there could be non trivial dependence on how the halo mass function falls off at the low mass end in different simulation. } The calibration was performed changing $\alpha_{\rm HOD}, \sigma_L , f$  while keeping certain parameters fixed, such as the radius up to which a halo is populated, which remains at \( 3R_{\text{vir}} \) to match with the \cite{2011ApJ...736...59Z} two point clustering for galaxies of different color types and luminosity bins. See \cref{app:calibration} for more details. For a more flexible calibration approach based on MCMC sampling, see Gonzalez et al., in prep.
    
    \item The number density of {\Halpha} emitters in the Flagship catalogue was calibrated to reproduce model 1 and model 3 of \cite{2016A&A...590A...3P}. Here, we do not change the original pipeline but rather see how well the original pipeline reproduces expected number counts from the Pozzetti model.
\end{itemize}
\subsection{Extension of the pipeline to z>3}
Below, we document our efforts of a preliminary extension of the pipeline up to $z = 10$. The spectral energy distribution (SED) assignment procedure in the default pipeline begins sampling luminosities from an evolving luminosity function (LF) constructed from a combination of the local SDSS LF \citep{2003ApJ...592..819B} and the CANDELS LF of \cite{2005ApJ...631..126D} defined up to $z\sim2$. To extrapolate this evolving LF to higher redshitfs, we forward model various forms for the luminosity function redshift evolution sampling absolute magnitudes, and using the existing pipeline to identify SEDs to convert luminosities into apparent magnitudes and compare them with the observed number counts in COSMOS2020 until a match is obtained(see \cref{fig:lumcalibration}).

This forward-modeling approach described above inherently involves assumptions regarding the SED of each galaxy. 
In order to minimize errors resulting from the resulting incorrect K-corrections when computing apparent magnitudes from luminosities, we adopt a strategy of comparing to redder observed bands while calibrating for higher redshifts. The default pipeline samples the rest-frame $r01$-band luminosity which corresponds to observed optical wavelength regions when sampling at low redshift. However, as the redshift increases, the light emitted in the rest-frame $r01$-band is progressively observed at redder wavelengths. To take this into account and maintain consistency with the original pipeline, we strategically shift to calibrating against successively redder photometric bands (e.g., ${\it i}_{\rm HSC}, {\it H}_{\rm UVISTA}, {\it ch1}\ _{\rm IRAC}, {\it ch2}\ _{\rm IRAC}$) as we move to higher redshifts (see \cref{fig:lumcalibration}). This ensures that we are always observing light that originates from the stellar main sequence-dominated and well-characterized rest-frame optical portion of the galaxy SED, rather than relying on extrapolations from the star formation-dominated highly variable and dust-affected rest-frame UV. By maintaining a more consistent sampling of the intrinsic SED across different redshifts, we aim to reduce the uncertainties associated with K-corrections and improve the accuracy of our derived photometric properties. 

Finally, the SED assignation pipeline is also modified to AM transform the {\it g01}-{\it r01} color distribution to distributions of redshifted colors in COSMOS2020.
Assigning SED to each galaxy enables us to compute apparent and absolute magnitudes in 30 different broadband filters ranging from UV to infrared in wavelength coverage.

\begin{figure}
\includegraphics[trim={0.2cm 0.2cm 0.2cm 0.2cm},clip,width=1\linewidth]{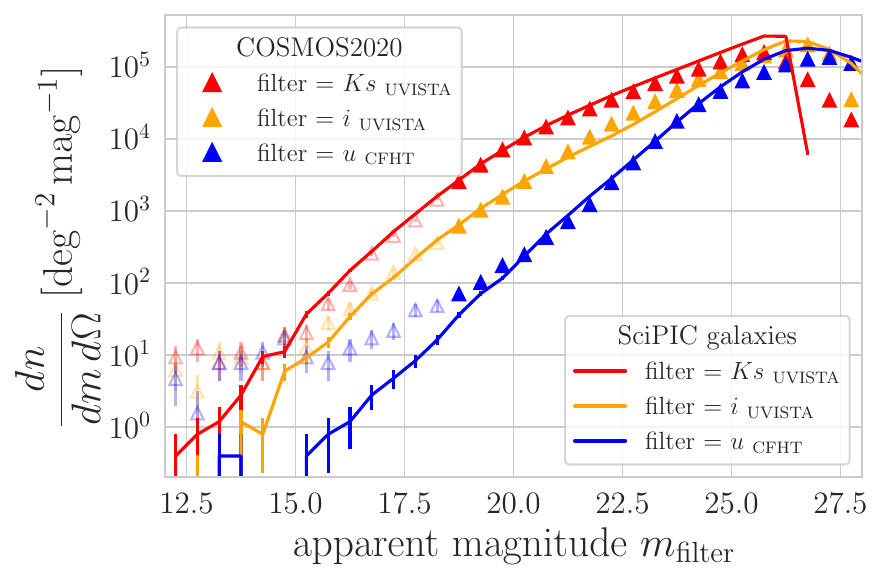} 
\caption{Number counts per unit apparent magnitude per unit solid angle from our mocks (solid lines) compared against COSMOS2020 (markers). The three different colors show the apparent magnitude counts in three different band filters ranging from UV to near infrared. The empty markers denote the unreliable range of COSMOS affected by saturation. }
\label{fig:numbercounts}
\end{figure}

\begin{figure}
\includegraphics[trim={0.2cm 0.2cm 0.2cm 0.2cm},clip,width=1\linewidth]{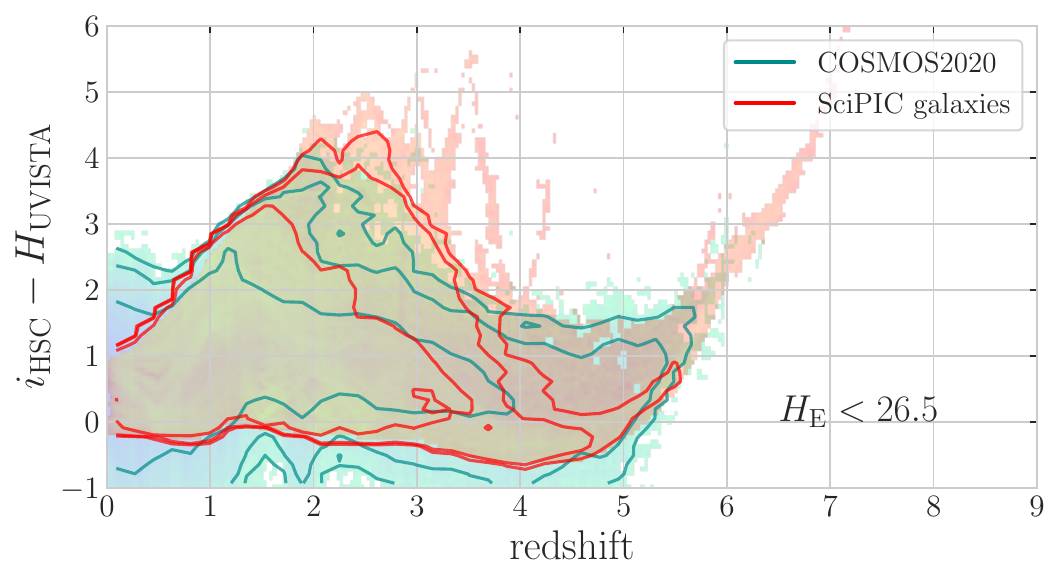} 
\caption{A comparison of the $i_{\rm HSC}-H_{\rm UVISTA}$ color distribution across redshift for COSMOS2020 vs our mocks. A magnitude cut of $m_{\rm NISP}<26.5$ has been placed on the mock galaxies.}
\label{fig:sed}
\end{figure}

In \cref{fig:numbercounts}, the solid lines show the resulting number counts 
in our mock galaxies 
compared to the markers which correspond to the number counts in COSMOS2020. The colors blue, orange, and red are used to denote magnitudes in  ${\it u}_{\rm CFHT}$, ${\it i}_{\rm UVISTA}$ and ${\it Ks}_{\rm UVISTA}$ filters, respectively. There is good agreement between the observations and the mocks at intermediate brightness. It is to be noted that the number counts from COSMOS2020 are unreliable at faint magnitudes due to incompleteness and at bright magnitudes because detector saturation and associated photometric systematics cause inaccurate fluxes and misclassification of bright sources. 
In \cref{fig:sed}, the $i_{\rm HSC}-H_{\rm UVISTA}$ color distribution is shown as a function of redshift in red for the mock galaxies and in teal for the distribution of COSMOS2020 galaxies. The drop-off in the number of objects in the color distribution at $z>6$ is due to the absorption of UV light due to the inter-galactic medium \citep{Madau1995} and the apparent magnitude selection cut applied. Throughout this section, we have chosen to apply magnitude cuts similar to the depth expected to be reached by the \Euclid's  NISP or Near Infrared Spectrometer and Photometer \citep{EuclidSkyNISP}. Specifically, we have applied a cut of ${\it H}_{\rm E}<26.5$ to reflect realistic observations similar to the Euclid Deep Fields. 

The galaxy catalogue produced in this work is further supplemented with derived physical properties, including stellar mass, star formation rate, metallicity, dust extinction, morphological parameters, intrinsic galaxy shapes, and emission-line measurements, all derived following the methodology of \citet{2025A&A...697A...5E} resulting in a total of 225 descriptive properties.

\subsection{ \Halpha selected tracers - number density and clustering}
\begin{figure}
    \centering
    \includegraphics[trim={0.2cm 0.2cm 0.2cm 0.2cm},clip,width=1\linewidth]{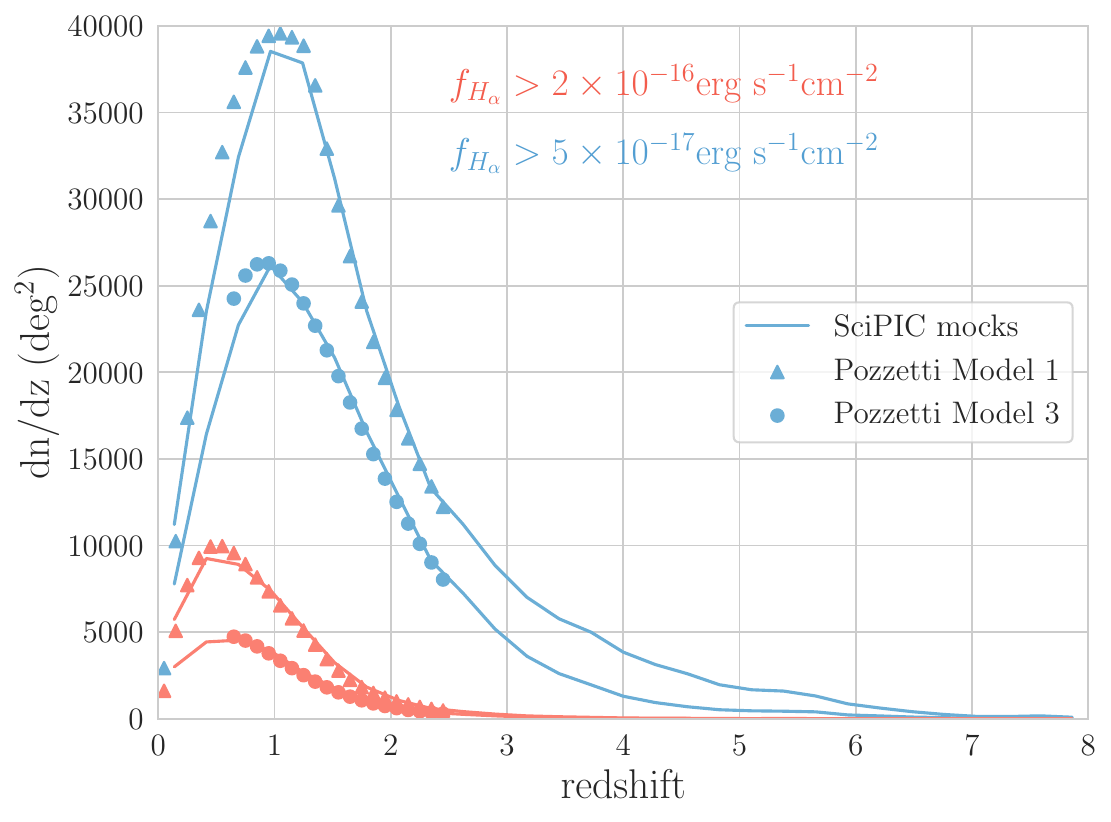}
\caption{Redshift distribution of \Halpha flux selected galaxies. The \cite{2016A&A...590A...3P} models (in markers) are compared with the \Halpha galaxies in our catalogue (solid lines). The red and blue colors are used to distinguish the galaxy counts with imposed flux cuts as shown in the label.}
    \label{fig:pozzetti}
\end{figure}

\begin{figure*}
\includegraphics[trim={0.2cm 0.2cm 0.2cm 0.2cm},clip,width=1\linewidth]{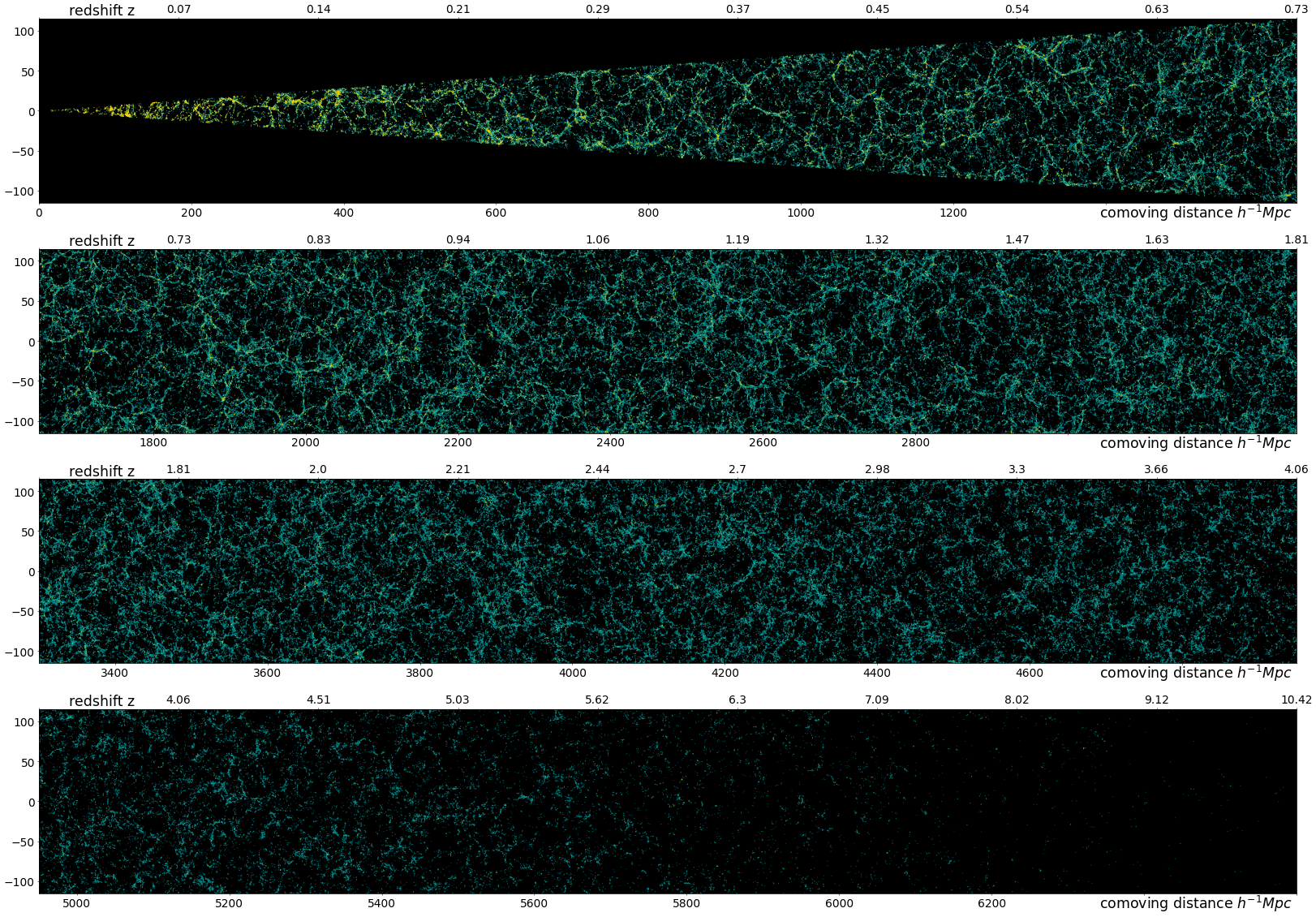} 
\caption{A $4\,{\rm Mpc}\,h^{-1}$  thick cross section of \Halpha flux selected galaxies in yellow painted on top of dark matter haloes in cyan from the Uchuu lightcone. The observer is placed at the origin, the length along the x-axis corresponds to line-of-sight direction of increasing comoving distance of the galaxies/haloes from the observer. A flux cut of $5\times 10^{-17} ~\mathrm{erg~s^{-1}~cm^{-2}}$ has been placed on the \Halpha emission line flux of the galaxies. A 3D interactive view is also available online at \url{https://rsujatha.github.io/lc-view/}.}
\label{fig:halpha}
\end{figure*}

In the following, we concentrate on \Halpha emission line galaxies that will become one of the main large scale structure tracers in future Stage IV surveys. The default pipeline outputs emission line fluxes for several elements.  
Here, we consider the \Halpha fluxes calibrated to two different \citeauthor{2016A&A...590A...3P} models and corrected for internal extinction. These \Halpha fluxes correspond to the \texttt{logf\_halpha\_model1\_ext} and \texttt{logf\_halpha\_model3\_ext} entries in the catalog.
The two models vary in their respective fits to the \Halpha luminosity function: model 1 fits a large range of observational data while model 3 focuses on fitting those data sets relevant for \Euclid and {\it WFIRST-AFTA} slitless surveys, i.e., for redshift ranges
0.7 < z < 2.23 \citep[see][for more details]{2016A&A...590A...3P}. \Cref{fig:pozzetti} shows a comparison of the redshift distributions of {\Halpha} selected samples on the Uchuu lightcone (solid lines) compared to the \citeauthor{2016A&A...590A...3P} models (in markers). 
The red lines and markers showcase the redshift distribution of the \Euclid Wide survey for its expected \Halpha flux limit of $2\times 10^{-16}\, \rm{erg}\,s^{-1}\rm{cm^{-2}}$.
The blue lines and markers correspond to an \Halpha flux cut of $5\times 10^{-17}\,\rm{erg}\,s^{-1}\rm{cm^{-2}}$, a forecast of what the \Euclid Deep fields will observe.
In \cref{fig:halpha}, we plot a thin slice of the \Halpha galaxies above a model 3 flux limit of $5\times 10^{-17} \rm{erg}\,s^{-1}\rm{cm^{-2}}$ on the Uchuu lightcone. The galaxies overall trace the cosmic web structure of the dark matter field as can be seen from the large-scale filamentary features, which are specially noticable at low redshift. However, at higher redshift the distribution of galaxies becomes more homogeneous reflecting less growth and a more linearly distributed matter distribution. Although we have populated the entire lightcone with \Halpha galaxies, the range of redshift relevant for \Euclid is only from (z=0.8 to 1.8) where the \Halpha emission line falls inside the wavelength range covered by the NISP instrument.


\begin{figure}
    \centering
    \includegraphics[trim={0.2cm 0.2cm 0.2cm 0.2cm},clip,width=1\linewidth]{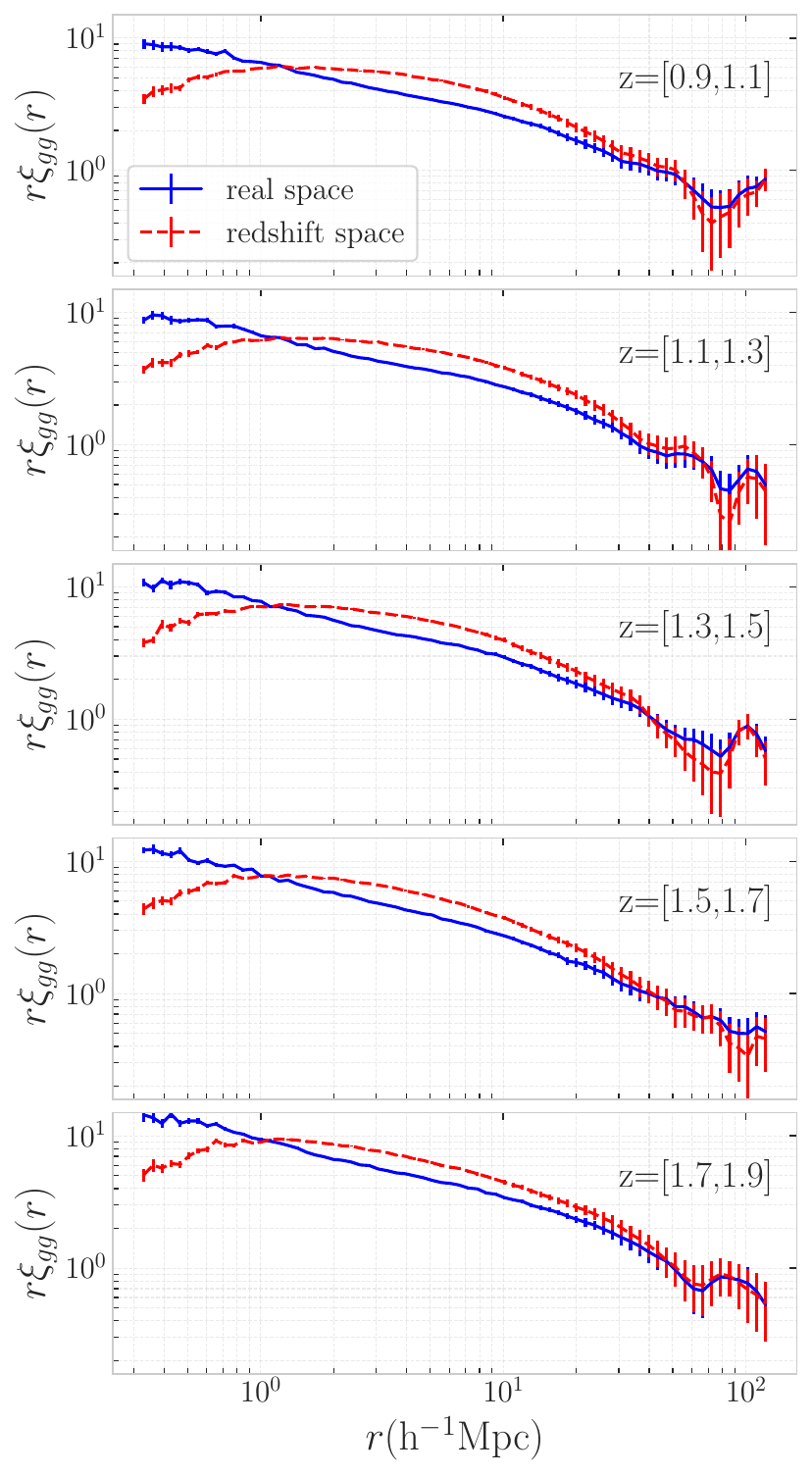}
    \caption{An estimate of the two-point correlation function in the \Halpha model 3 galaxy catalog. The redshift-space correlation function is shown in red while the real space clustering is shown in blue. The panels from top to bottom show the results in increasing redshift bins. The $50\,{\rm deg}^2 $ circular patch is divided into 32 cutskies used to
estimate jackknife errors.}

    \label{fig:predict}
\end{figure}

One of the main statistics used for cosmological inference in surveys is the two-point correlation function of the galaxies. It is a statistical measure that quantifies the excess probability of finding a pair of galaxies at a given separation compared to a random distribution. We show the two-point galaxy-galaxy correlation function for the \Halpha flux selected galaxy sample mentioned above in \cref{fig:predict}.
We compute the two-point correlation function in 70 linear bins of separation distance from  $0.3\,\hMpc$ to  $120\,\hMpc$. We show both the real and redshift-space correlation functions in blue and red, respectively, in all panels. As expected the redshift-space clustering at small scales is suppressed compared to the real space clustering due to random motion inside galaxy clusters and it is enhanced at larger scales due to the coherent infall of galaxies into high-density gravitational potentials \citep{1987MNRAS.227....1K,2001Natur.410..169P}. The different panels present the two-point correlation at different redshifts, showing a small increasing trend in the correlation function power with increasing redshift. \Cref{sec:LSestimate_validation} details our procedure to compute the correlation function. In some of the panels, there are hints of the presence of the Baryon Acoustic Oscillation (BAO) peak. 
\section{Summary and Discussion}
\label{sec:sumamry}
This paper introduces \holcon, a toolkit for interpolating haloes and producing lightcones and demonstrate its use for large scale cosmological surveys by generating mock galaxy catalog. \holcon includes a module that identifies optimal angles within the simulation box to orient the lightcone to avoid repeating structures along the line of sight. While previous methods \citep{2009A&A...499...31H,2010ApJS..190..311C} have focussed on non repetitive volumes that are box-like and elongated, the method derived in \cref{sec:ooa} opimises for conical geometry with tunable parameters relavant for a sky survey, such as the depth and angular radius of the footprint. \Cref{fig_sym} shows a diagramatic representation of the lightcone of a given length and angular radius. Then we formulate the concept of the `repetition length' for any direction. \Cref{fig_lattice} is a 2D lattice, with repetition length shown in three different directions for demonstration purposes. Using this formalism we first analytically derive a mask function $M(\alpha,\delta)$ in \cref{eq:main} that takes a value > 0 for orientations $(\alpha,\delta)$ where there is no repetition for the given geometry of the lightcone.
For different combinations of the length and angular radius of the lightcone we plot this mask function in \cref{fig:config1}. In the same figure, we also shows a comparison of the mask function to identify unique structures from \cref{eq:main} on the right panel against a numerical search for fraction of unique structures on the left panel. For further insight, \cref{fig:numerical} is used to identify the lattice points of a cube overlapping with the on a map showing the fraction of unique volume in a lightcone along any direction.

In addition this, \holcon also provides a module to linearly interpolate haloes between snapshots in order to identify and extract those haloes that cross the lightcone. This module is compatible with a the \dask distributed computing framework, enabling fast halo matching between snapshots using relational join operations. \Cref{tab:join_operations} describes the halo IDS used to track merger history from two different halo finders and merger tree  algorithms namely \texttt{ROCKSTAR+CONSISTENT TREES} and \texttt{HBT-HERONS}. In \cref{sec:gal}, we use this module to construct a 50 $\rm{deg^2}$ lightcone with a depth up to $z = 10$ using the publicly available Uchuu simulations. This halo lightcone is to provide representative coverage to the deep fields of Stage IV surveys. \Cref{fig:schematic} shows a schematic representation of the lightcone with the different snapshots shown as thin color-coded shells placed exactly at the halo crossing location. The choice of the specific orientation of the lightcone is done taking into account the identification of the optimal angles developed in \cref{sec:ooa}. We perform validation tests to check its utility for cosmological purposes. 
    We have compared the HMF in the lightcone as a function of redshift to the corresponding
     measurement in the simulation snapshots. As illustrated in \cref{fig:massfn}, the cumulative halo mass functions show good agreement, and the relative errors are not greater than 20\% in the relevant mass ranges, thus providing a validation of the lightcone construction.
\Cref{fig:massfn} presents a comparison of the two-point correlation function of halo mass selected samples measured in both configuration and redshift-space in the lightcone and in the snapshots at various redshifts. Again, the measurements in lightcone redshift shells show good agreement with those at similar fixed redshift in the snapshots, further validating our lightcone construction.

Finally, we create a mock galaxy catalogue by running the {\scipic} pipeline on the Uchuu lightcone %to generate a mock galaxy catalog
and compute several distributions of the galaxy observable properties. \Cref{fig:mass-lum-relation} shows the abundance matched relation between the halo mass and luminosity which is used to assign galaxies a value for their $r01$-band luminosity. We also generate an \Halpha flux-selected galaxy catalogue that mimics the selection of Stage IV surveys. 
We perform several validation tests to compare the catalogue against observations. 
\Cref{fig:numbercounts} compares the number counts per unit apparent magnitude per unit solid angle in the Uchuu mocks against COSMOS2020 observations. \Cref{fig:sed} compares the  $i_{\rm{HSC}}-H_{UVISTA}$ color as a function of redshift for the Uchuu mocks against the COSMOS2020 observations. To mimic a realistic survey, we perform flux cuts on the general catalogue to generate a \Halpha tracer population based on two different models. \Cref{fig:pozzetti} compares the redshift distribution of the \citeauthor{2016A&A...590A...3P} models against the mocks for \Halpha flux-limited samples. 
\Cref{fig:halpha} shows the visualization of a thin slice of the \Halpha flux-selected galaxies in the lightcone.
\Cref{fig:predict} shows the two-point correlation function clustering of the \Halpha flux-selected galaxies. 

We note that the analysis of galaxy clustering data binned by redshift may not be susceptible to systematic effects arising from the choice between repetitive and non-repetitive viewing angles. Effects from repetition along the line of sight are more relevant in observables integrating multiple redshifts such as weak gravitational lensing observables. For example, \citet{2024MNRAS.534.1205C} shows that the convergence power spectrum and higher-order moments of lensing fields can accumulate systematic effects depending on the viewing angles for small sky coverage.
Although we have chosen to generate an \Halpha flux-selected sample to showcase the capabilities and properties of our lightcone and galaxy generation pipelines, the Uchuu lightcone constructed in this work is general-purpose and can be used to construct representative catalogues expected to be observed by other deep surveys — for example, from radio-selected HI surveys of deep SKA fields \citep{2020MNRAS.499.3434Z} to rest-frame UV-selected Lyman $\alpha$ emitters from surveys such as ODIN \citep{2024ApJ...962...36L}. The pipeline has been implemented with a single HOD model and parameter set, but is intended to retrieve different tracer populations naturally by imposing appropriate flux cuts and instrument-specific systematics. In future work, we will revisit this framework with alternative tracer populations.
\subsection{Data availability}
The halo lightcone and galaxy catalogues can be accessed from the CosmoHub \citep{2020A&C....3200391T,2017ehep.confE.488C} platform after registering at \url{https://cosmohub.pic.es}. 

\bibliographystyle{aa} % style aa.bst
 \bibliography{references.bib} % your references Yourfile.bib
 \begin{acknowledgements}
SR thanks Aseem Paranjape, Violeta Gonzalez Perez, Mar Mezcua, Małgorzata Siudek for useful discussions.
SR also thanks Manuel Ruiz-Herrera, Alejandro Eróstegui Losantos for help with various concepts. 
SR acknowledges support from project “Advanced Technologies for the exploration of the Universe”, part of Complementary 
Plan ASTRO-1390 HEP, funded by the European Union - Next Generation (MCIU/PRTR-C17.I1). FJC and PF acknowledge support form the Spanish Ministerio de Ciencia, Innovación y Universidades,
MICIU/AEI /10.13039/501100011033,
projects PID2019-11317GB,

PID2022-141079NB, PID2022-138896NB; the European Research Executive

Agency HORIZON-MSCA-2021-SE-01 Research and Innovation programme
under the Marie Skłodowska-Curie grant agreement number 101086388 (LACEGAL) and the programme Unidad de Excelencia María de Maeztu, project

CEX2020-001058-M. JC and EG acknowledges support from the grant PID2021-123012NA-C44 funded by MCIN/AEI/ 10.13039/501100011033 and by “ERDF A way of making Europe”. EG also acknowledges grant from the European Union NextGenerationEU(PRTR-C17.I1) and by Generalitat de Catalunya. ME acknowledges support from the Spanish
Ministry of Science and Innovation through the project PID2023-152069NA-I00. ZB is supported by the Spanish Ministry of Science, Innovation and Universities through a doctoral program associated with the Unidad de Excelencia “María de Maeztu” (CEX2020-001058-M). The data production, processing, and analysis tools for this paper have been developed, implemented and operated in collaboration with the Port d’Informació Científica (PIC) data center. PIC is maintained through a collaboration agreement between the Institut de Física d’Altes Energies (IFAE) and the Centro de Investigaciones Energéticas, Medioambientales y Tecnológicas (CIEMAT).

We acknowledge the PIC (Port d’Informació Científica) data center for providing computing resources and support. This work has also made use of CosmoHub, developed by PIC (maintained by IFAE and CIEMAT) in collaboration with ICE-CSIC. It received funding from the Spanish government (grant EQC2021-007479-P funded by MCIN/AEI/10.13039/501100011033), the EU NextGeneration/PRTR (PRTR-C17.I1), and the Generalitat de Catalunya. We thank Instituto de Astrofisica de Andalucia (IAA-CSIC), Centro de Supercomputacion de Galicia (CESGA) and the Spanish academic and research network (RedIRIS) in Spain for hosting Uchuu DR1, DR2 and DR3 in the Skies \& Universes site for cosmological simulations. \AckECon
\end{acknowledgements}

\begin{appendix}

\section{Numerical search algorithm for the fraction of repeated volume}
\label{app:numerical}
\begin{figure}[hbt!]
\centering
\includegraphics[width=0.95\linewidth]{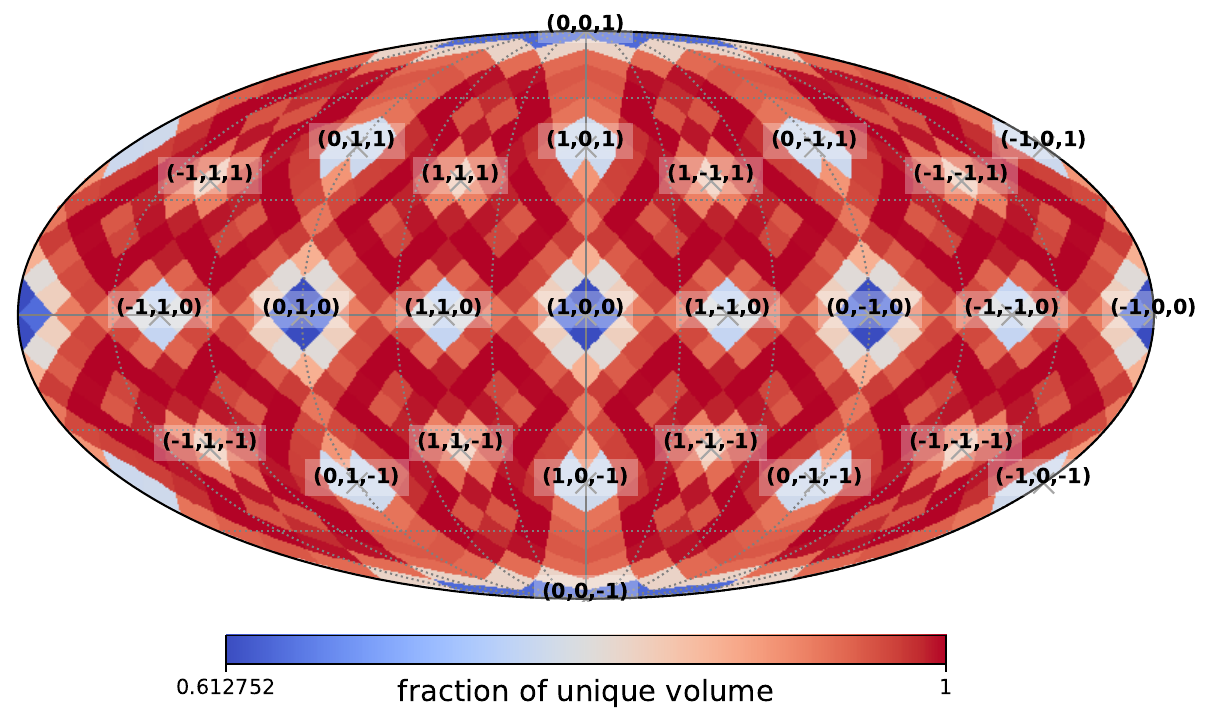}
\caption{The result of the numerical algorithm that identifies the fraction of unique volume probed when the lightcone is oriented along various directions. The regions where repetition occurred are colored in shades of blue and can be identified to be directed at the corners of the periodic boxes and marked with lattice point coordinates. The color scheme closely follow that of Figure 7 of \citet{2024MNRAS.534.1205C} which are plots generated by their numerical code (\textsc{line-of-sight finder}) so as to check the consistency with their results. This numerical algorithm is primarily used to compare against the analytical mask introduced in our work i.e., compare left and right panels of \cref{fig:config1}.}
\label{fig:numerical}
\end{figure}
In this appendix, we do a brute force computation to find the fraction of volume that is unique for different orientation angles in a lightcone of circular footprint and sky coverage of $50\, \mathrm{deg}^{2}$. A fraction value of one indicates unique structures along that direction.

A grid of points are generated inside the cubic box and labelled with a unique \textsc{id}. This set-up of a grid of points with a unique \textsc{id} represents a periodic cosmological box with unique cosmic-web structures. The cubic box is also repeated on all sides with grid points labeled with the same unique \textsc{id} as the primary box. 
In the example in \cref{fig:numerical}, the primary box is chosen to be of $20 \, \hMpc$ length and the length of the lightcone is $75 \, \hMpc$ making their length ratio $l/L_{\rm{Box}} = 3.75$. We define the declination angle with respect to the angle made with the z axis which is oriented along the cube side. The colors in \cref{fig:numerical} indicate the fraction of volume or, equivalently, the fraction of \textsc{id} that are unique. We can identify that the minima of the map or those with most repeated structures are the directions corresponding to the edges, the face diagonal, and the body diagonal of the periodic simulation box. Subsequent local minima or lighter shades of blue occur at the edges, face diagonal body diagonal of the adjacent repeated boxes. \Cref{eq:main} 
%of this work 
is a threshold function corresponding to the map in \cref{fig:numerical}. It takes a value of 1 when the value of the map or fraction of unique volume is 1 and 0 otherwise.

\section{Validation of the lightcone for cosmological inference.}
\label{app:validation}
\begin{figure}
\includegraphics[width=\linewidth]{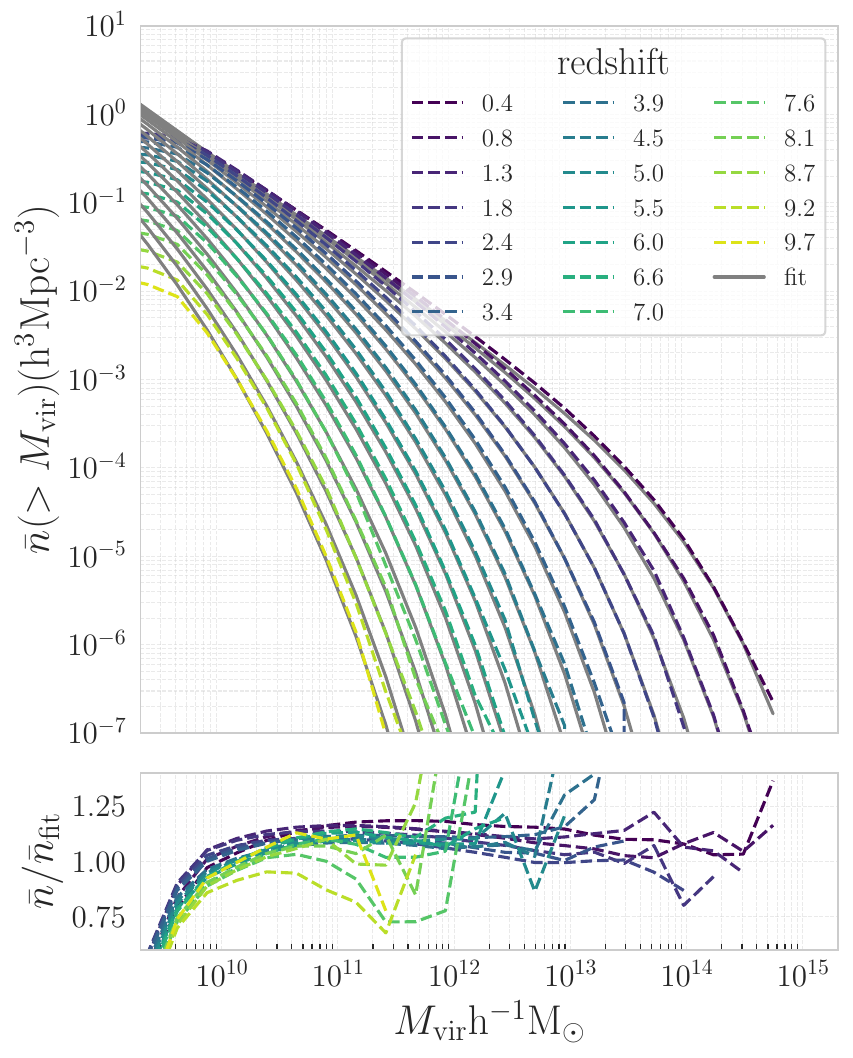}
\caption{\textbf{Validation of halo mass function of the Uchuu lightcone}: The dashed lines in the top panel shows the cumulative halo mass function in narrow sections of the lightcone as indicated by the mean redshift in the labels. The solid grey lines that closely match the dashed lines show a comparison with a 5 parameter Schechter-like function fit to the cumulative halo mass function from the simulation snapshots. \emph{Bottom}: Ratio of the cumulative HMF computed from simulations to those from the fits.}
    \label{fig:massfn}
\end{figure}

\begin{figure*}
    \includegraphics[trim={0.25cm 0.3cm 0.25cm 0.25cm},clip,width=1\linewidth]{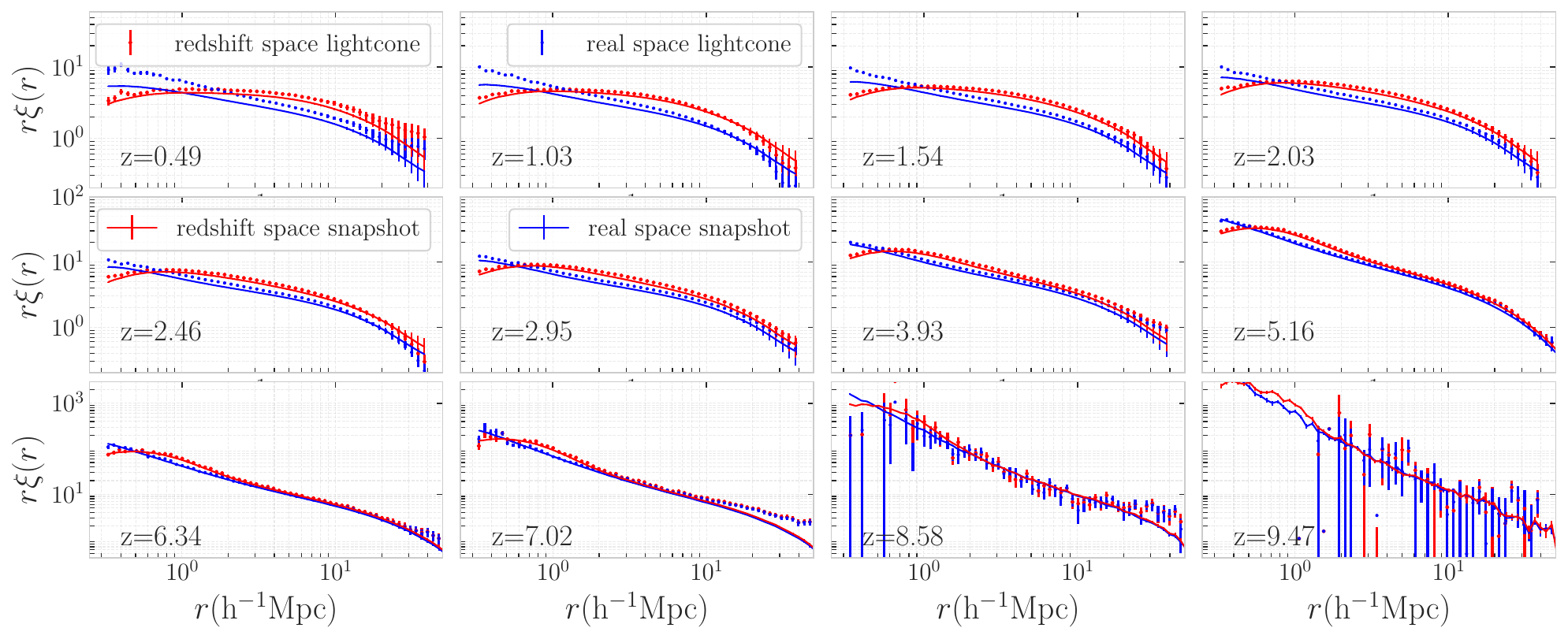}
    \caption{\textbf{Validation of the two-point correlation function of the haloes in the Uchuu lightcone}. A comparison of the two-point correlation function in the lightcone (markers) against those computed at constant redshift (lines) from the snapshots. A total of 50 logarithmic bins between separation distance 0.3 to 38 $\,h^{-1}\,{\rm Mpc}$ have been used to compute the correlation function. The redshifts of comparison vary across different panels, starting from $z=0.49$ at the top left panel to $z=9.47$ at the bottom right panel. In the lightcone, the correlations are computed in thin slices of thickness $\log_{10} \Delta z = 0.04$ centered in the redshift of interest. Both the real and redshift-space clustering are plotted to assess our position and velocity interpolation schemes. The error bars are computed using jacknife estimates. We find that there is a fair agreement between all curves at intermediate redshifts, validating our construction pipeline. At the extreme redshifts, the mismatches seen in the correlation functions are due to cosmic variance arising from small volume at the low redshift end and small halo number densities at the high redshift end.}
    \label{fig:2ptcorr}
\end{figure*}

%To validate the utility of the lightcone, in this section, 
To validate the lightcone generation procedure, 
we compare the one-point and two-point statistics from the lightcone to those from the simulation snapshots. \Cref{fig:massfn} shows a comparison of the halo mass function. The lightcone is divided into 20 bins in redshift/comoving distance. In the top panel, we plot the cumulative halo mass function  within each of the redshift bins with dashed lines. The different colors correspond to the mean redshift within those redshift bins Since we have snapshots available only at discrete redshift intervals, we first fit the mass function to a 5-parameter Schechter-like function\footnote{We define a 5-parameter Schechter-like function adding a linear dependence of the power-law exponent and a power-law dependence to the term in the exponential factor of the 3-parameter Schechter function. The resulting function is $f(x) = N\, \left(\frac{x}{x_*}\right)^{\left[\alpha_1 + \alpha_2\,\log_{10}\left(\frac{x}{x_*}\right)\right]}\, \exp\left[-\left(\frac{x}{x_*}\right)^\beta \right] $, where $N$, $x_*$, $\alpha_1$, $\alpha_2$, and, $\beta$ are the five parameters of the function.} with redshift evolution in the five parameters. The fit is performed only for haloes with masses above $10^{11} h^{-1}M_{\odot}$ to prevent fitting for the spurious drop in the mass function at low masses. This fit is then plotted in grey solid lines in \Cref{fig:massfn} to facilitate the comparison with the mass function from the lightcone. The dashed lines in the bottom panel show the ratio of the two mass functions. We can see that at very low masses the lightcone has an underprediction of haloes due to resolution effects. At very high masses the noisy curve is due to few number of haloes in our sample. At intermediate masses, there is a good match between the lightcone and simulation boxes halo mass functions, thus validating the halo mass function of the lightcone. 

\Cref{fig:2ptcorr} shows the result of a comparison of the two-point correlation function from the lightcone with the simulation snapshot. Here, we select all the haloes above $10^{11} h^{-1}M_{\odot}$ to ensure well-defined masses. To test/verify that both positions and velocities have been interpolated accurately, we compute and compare both the real and redshift-space two-point correlation functions. \Cref{fig:2ptcorr} illustrates these comparisons, displaying the two-point correlation function $r \xi(r)$ as a function of separation ($r$) for both real space in blue and redshift-space in red. At small scales, the expected correlation function in redshift-space is suppressed due to random motions, while at large scales the correlation function has a larger amplitude than the real space correlation function due to coherent infall into the centers of the overdensities. The different panels in \cref{fig:2ptcorr} repeat the analysis for twelve different redshifts from $z=0.49$ to $z=9.47$.
The lightcone is divided into slices of $\Delta \log_{10} z=0.04$ width
centered at the redshifts of the twelve chosen snapshots.
Except for the largest and the smallest redshifts where there are limited number of haloes in a $50$ $\rm{ deg}^2$ patch, the two-point correlation functions from the lightcone, denoted by the markers, agree well with the analysis from the individual snapshots, denoted by solid lines, thus validating the construction pipeline.

\section{Calibration of \scipic}
\subsection{Halo mass - incompleteness and discreteness correction}
\label{app:masscorrection}
\begin{figure}
    \centering
    \includegraphics[width=0.98\linewidth]{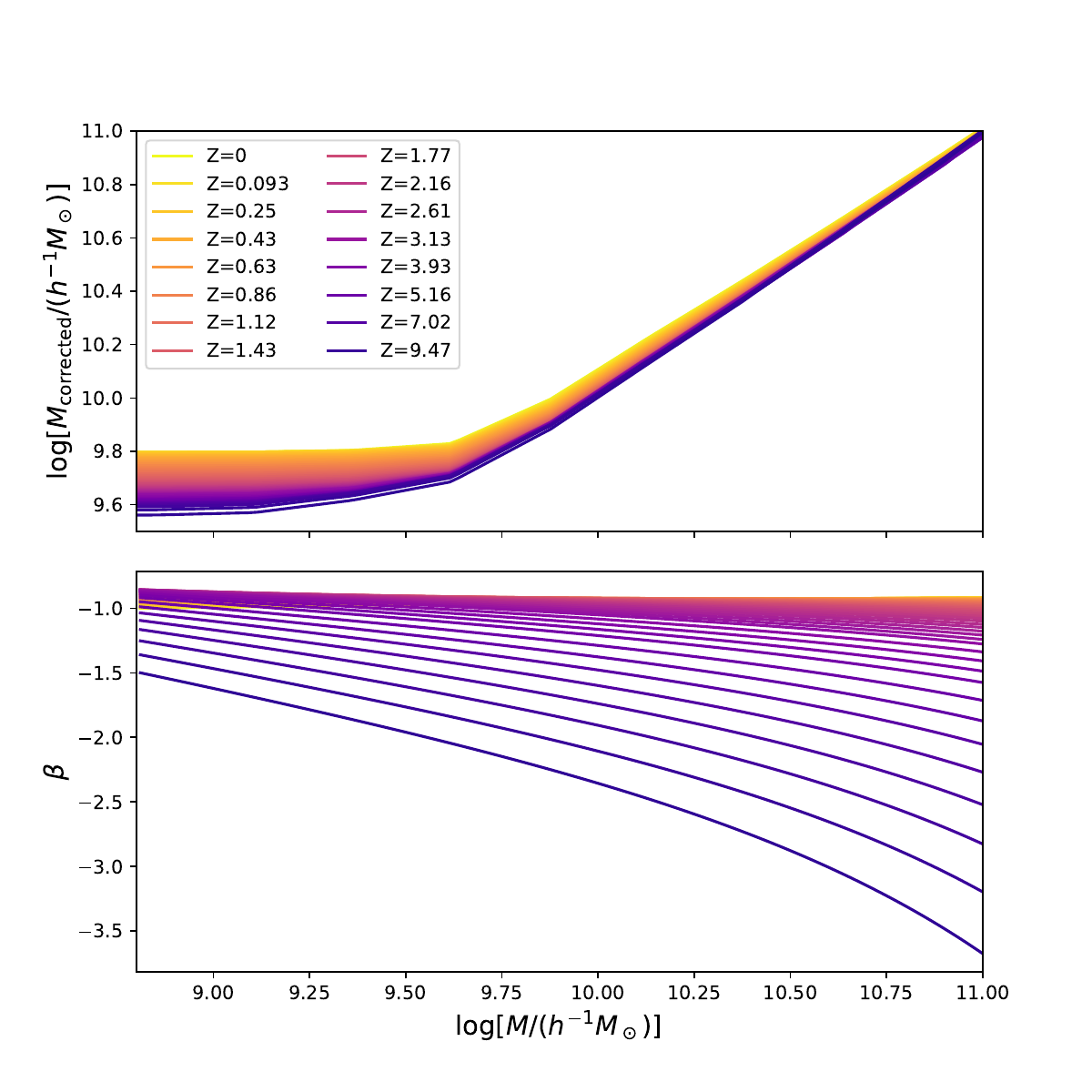}
    \caption{\emph{Top}: The monotonic relation between the original halo mass and corrected halo mass obtained from abundance matching the non-convergent measured halo mass function in Uchuu with corresponding HMF fits (dashed lines with soild lines in \cref{fig:massfn}). \emph{Bottom}: The slope $\beta$ as a function of mass while describing the HMF as a power law. We use this to correct for the discreteness in HMF. In both panels, the different colors correspond to the differnt redshift as indicated in the legend.}
    \label{fig:masscorrections}
\end{figure}

Surveys that probe to great depth aim to resolve fainter objects at low redshift and the brightest objects at high redshift, which preferentially occupy low-mass and high-mass halos, respectively. Consequently, such deep surveys naturally require simulations with high dynamic range, often pushing computational limits. However, due to the finite resolution of any simulation, the resulting halo mass function exhibits a spurious drop in the number of halos at the low-mass end—a well-known issue, particularly for mock survey catalogs. Several studies have attempted to mitigate this problem \citet{2014MNRAS.442.3256A,2022MNRAS.510L..29A}. Here, we follow a practical approach of abundance matching the cumulative HMF computed from the Uchuu simulation with the corresponding 5 parameter Schechter-like function fit. These cumulative functions correspond to dashed lines and solid lines respectively in \cref{fig:massfn}. The resultant monotonic transformation from original masses to corrected masses shown in the top panel of \cref{fig:masscorrections} ensures completeness of the HMF. 

At extremely low masses, haloes are comprised of few dark matter particles making the HMF discrete instead of a continuous function. By assuming the cumulative HMF (CHMF) to follow a power law with mass, i.e, CHMF $\propto M_{\rm h}^{\beta}$, we can sample a uniform variate $u$ and reassign the halo masses in the following,
\begin{equation}
    M_{\rm h}^{\prime} = [u(M_{h}^{\beta}-(M_{h}+m_{\rm p})^{\beta})+(M_{\rm h}+m_{\rm p})^{\beta}]^{1/\beta}
\end{equation}
where $M_{\rm h}^{\prime}$ is the reassigned mass that corrects for discreteness in the HMF. The bottom panel in \cref{fig:masscorrections} shows the fits for the slope $\beta$ in the Uchuu cumulative halo mass function which is required for the discreteness correction in the above equation.
\subsection{Luminosity and color dependent clustering}
\label{app:calibration}

\begin{figure*}
%\sidecaption
\includegraphics[width=0.9\textwidth]{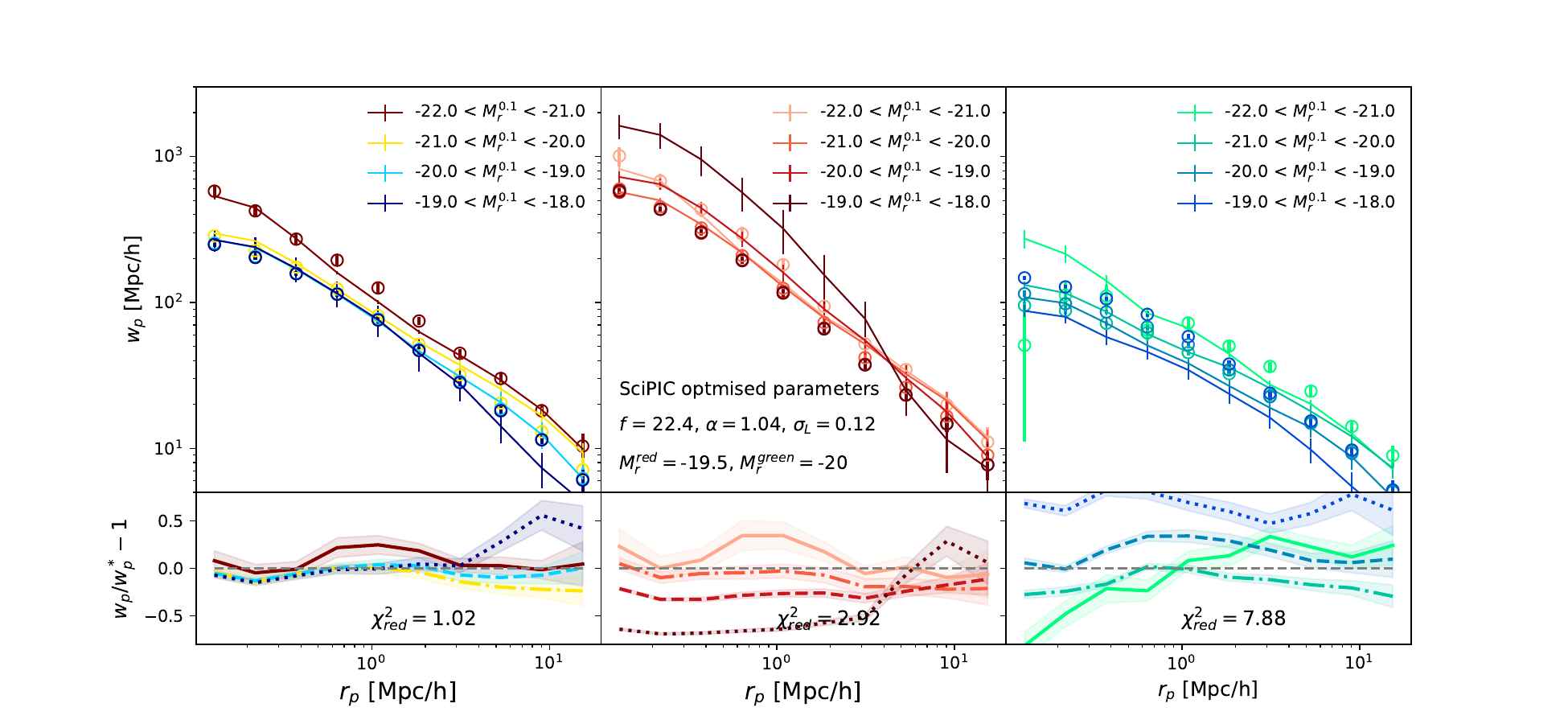}
\caption{A comparison of the projected two-point correlation function for different luminosity bins in the mocks (dots) and the observations (solid lines). A total of 10 logarithmic bins from 0.1 to 20 ${\, h^{-1}\,\rm Mpc} $ have been used for computing the projected correlation function. The optimised HOD and AM parameters are $f = 22.4$, $\alpha_{\rm HOD} = 1.04$, and $\sigma_{L} = 0.12$. \emph{Left}: The colours indicate galaxy populations in $r01$-band luminosity bins, \emph{Central}: Red galaxies in different luminoity bins, \emph{Right}: Blue galaxy population in different luminosity bins. } 
\label{fig:calibration}
\end{figure*}
To calibrate the parameters, we have selected a cubic subvolume of length $300\, \hMpc$ and repopulate it with different set of parameters to identify the set of parameters that generate a catalogue that better reproduce the two-point projected correlation function of \citet{2011ApJ...736...59Z}. The set of parameters that are varied are the HOD parameters, i.e., $f,\alpha_{\rm HOD}$ from \cref{eq:satellite}, the AM scatter parameter $\sigma_{L}$ and the parameters that go into modelling the fraction of red, blue and green galaxies at any luminosity $M^{\rm red}_{r}$ and $M^{\rm green}_{r}$. To model the luminosity dependence of clustering, 

the parameter calibration is performed by diving the galaxy catalogue into four luminosity bins (left panel in \cref{fig:calibration}). 
%We also further subdivide the galaxies according to the red or blue type and take their luminosity dependence with clustering in the middle and right panels.
We also calibrate the clustering color dependence by further subdividing the galaxies into red and blue types (middle and right panels in \cref{fig:calibration}).

\subsection{$r01-$ band luminosity function}
\begin{figure}
%\sidecaption
\includegraphics[width=0.9\linewidth]{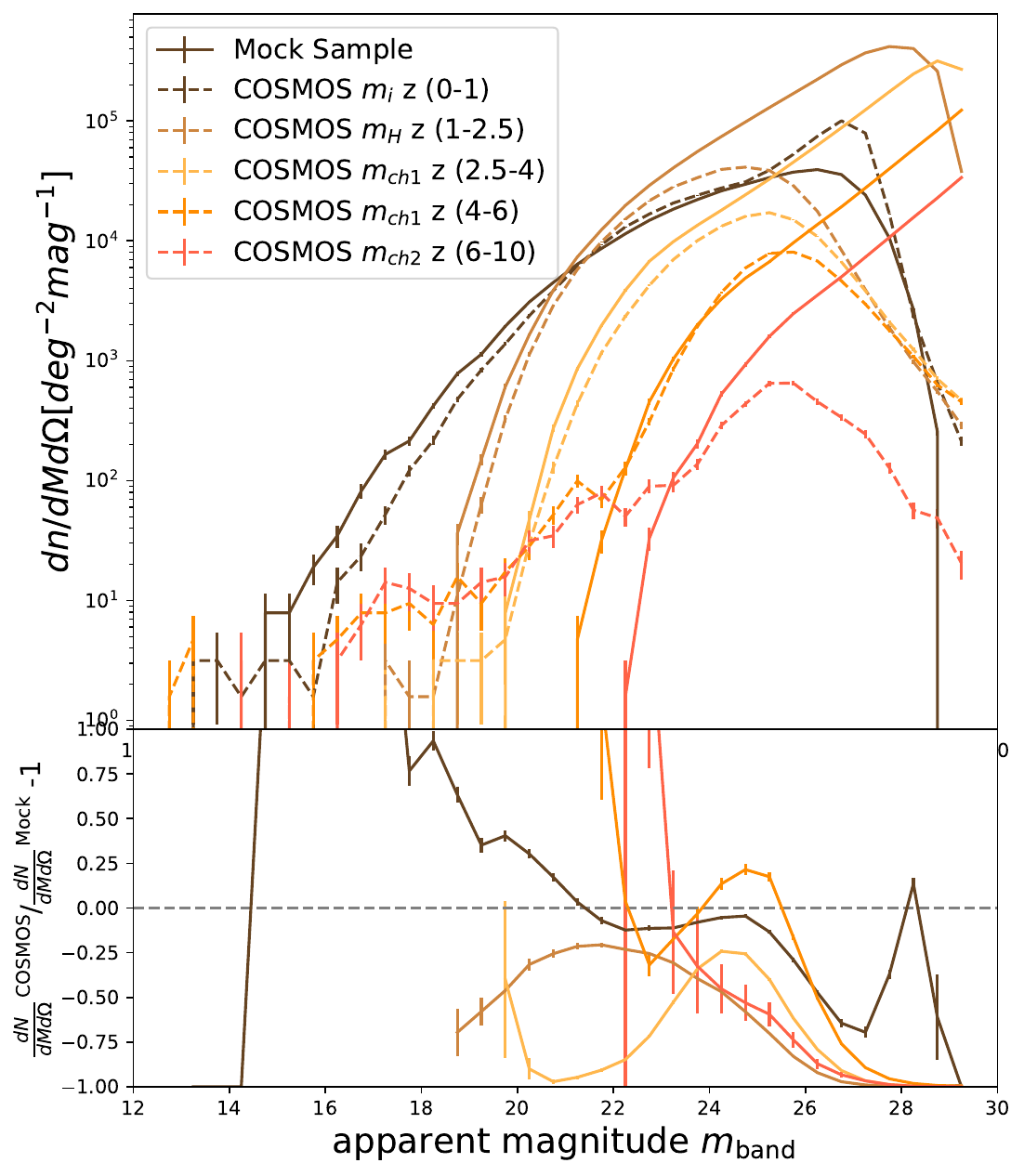}
\caption{\emph{Top}: A comparison of the apparent magnitude number counts using a forward modelled mock observation (solid lines) and COSMOS2020 (dashed lines). The colors from dark to light tone denote increasing redshift bins and apparent magnitude distributions in bands of increasing wavelength. \emph{Bottom:}The fractional error in the number counts in mock compared to the observations. } 
\label{fig:lumcalibration}
\end{figure}

In our galaxy mock the $r01-$band luminosity function has been sampled using the luminosity function fits from \citet{2003ApJ...592..819B} and \cite{2005ApJ...631..126D}. These LFs are based on observations from SDSS and CANDELS and reproduce observed data for redshifts $z<2$. To extrapolate the LF to higher redshifts, we consider several functional forms for the %decrement
evolution of the luminosity function for higher reshifts 
and create samples or realisations of galaxies with those luminosity distributions in a lightcone. We then assign to each galaxy an SED based on our pipeline, and compute their magnitudes in different filters applying K-corrections. We obtain a distribution of apparent magnitudes that can be compared against the deep number counts observations of COSMOS2020 to identify the best fitting %functional decrement
evolution of the luminosity function. Ultimately, we adopted the following modification to the $r01-$band luminosity function implemented in the default version as shown in \cite{2025A&A...697A...5E}
\begin{equation}
    \phi = \phi_{*}\, \left(\dfrac{3}{1+z}\right)^{3} \hspace{40pt} {\rm for }\hspace{5pt}z>2 \,,
    \label{eq:evolution}
\end{equation}
where $\phi$ is the new normalisation of the Schechter luminosity function, and $\phi_{*}$ the normalisation in the default pipeline.
Since the rest-frame $r01$-band progressively moves to redder observed bands for high redshift galaxies sampling,
%Since the part of the SED probed by $r01$ band progressively moves to redder bands for high redshift galaxies, 
we have used apparent magnitudes from $i_{\rm HSC}, H_{\rm UVISTA}, ch1\ _{\rm IRAC},ch2\ _{\rm IRAC}$ 
%for increasing redshift bins 
in order to sample the same rest-frame $r01$-band at different redshifts and converge on \cref{eq:evolution} for the redshift evolution of the luminosity function. 
\Cref{fig:lumcalibration} illustrates the calibration procedure. Darker shades represent the comparison between the apparent magnitude number counts of mock galaxies and observations at low redshifts in the lower-wavelength filters, while lighter shades show the comparison at higher redshifts using the higher-wavelength filters.

\section{Computation of the two-point correlation function}
\label{sec:LSestimate_validation}
We first discuss the procedure to obtain the real and redshift-space two-point correlation functions for both galaxies and dark matter haloes. The real space correlation can be just obtained by computing the clustering of the comoving positions of the haloes or galaxies. In redshift-space the location of the object appears displaced along the line-of-sight. This additional displacement can be obtained by adding to the z direction, a factor of $v_{z}/aH$, and to the lightcone adding  a term $v_{\rm los}/(aH)$ to the comoving z direction and a term $v_{\rm los}/(aH)$ to the line-of-sight comoving distance.  
We have used the Landy Szalay estimator \citep{1993ApJ...412...64L} to compute the correlation function, i.e.,
\begin{equation}
\xi(r) = \dfrac{DD-2DR-RR}{RR}\,,
\end{equation}
where $DD$ is the total number of pair counts of galaxies/haloes with separation between $r$ and $r+dr$, $RR$ is the pair count of a random distribution of points in the same region and $DR$ is the pair count of the data points with the random points. 
The $50\,{\rm deg}^2$ circular patch is divided into 32 cutskies to estimate jackknife errors on the correlation function, while the boxes are divided into 64 rectangular jackknife regions. We used theoretical randoms for the snapshot boxes and created randoms within a shell geometry corresponding to the above slices. The random catalogues are generated to have a density 10 times larger than the halo number density to improve the signal-to-noise. All jackknife regions, errors, and two-point correlation function estimates were computed using \texttt{pycorr} \citep{2024ascl.soft03009D}.

\end{appendix}

\end{document}